\documentclass[12pt,draftcls,onecolumn,journal]{IEEEtran}
\usepackage{amsfonts}
\usepackage{amsfonts}
\usepackage{setspace}

\usepackage{amsmath}
\usepackage{amssymb}
\usepackage[dvips]{graphicx}
\usepackage{epsfig}

\newcommand{\tsnr}{{\text{\footnotesize{SNR}}}}
\newcommand{\tmin}{\text{min}}
\newcommand{\E}{\mathbb{E}}
\newcommand{\e}{\mathcal{E}}

\newcommand{\R}{{\sf{R}}}
\newcommand{\Pb}{\bar{P}}
\newcommand{\sta}{{\alpha^{\ast}_{\text{opt}}}}
\newcommand{\alphao}{\alpha_{\text{opt}}}
\newcommand{\ro}{r_{\text{opt}}}
\newcommand{\stro}{{{r^{\ast}_{\text{opt}}}}}
\newcommand{\dao}{\dot{\alpha}_{\text{opt}}}
\newcommand{\dro}{\dot{r}_{\text{opt}}}
\newcommand{\figsize}{0.65}
\newcommand{\rhoo}{\rho_{\text{opt}}}
\newcommand{\tsnref}{\tsnr_{\text{eff}}}
\newcommand{\tsnrefo}{\tsnr_{\text{eff,opt}}}

\newtheorem{Lem}{Theorem}

\newtheorem{Corr}{Corollary}

\begin{document}

%
\title{The Impact of QoS Constraints on the Energy Efficiency of Fixed-Rate Wireless Transmissions}



%

\author{\vspace{0.3cm}
\authorblockN{Deli Qiao, Mustafa Cenk Gursoy, and Senem
Velipasalar}
\thanks{The authors are with the Department of Electrical
Engineering, University of Nebraska-Lincoln, Lincoln, NE, 68588
(e-mails: qdl726@bigred.unl.edu, gursoy@engr.unl.edu, velipasa@engr.unl.edu).}
\thanks{This work was supported by the National Science Foundation under Grants CCF -- 0546384 (CAREER) and CNS -- 0834753.}}
%
%
%
\maketitle
\thispagestyle{empty}
\vspace{-0.8cm}
\begin{abstract}
Transmission over wireless fading channels under quality of service
(QoS) constraints is studied when only the receiver has channel side
information. Being unaware of the channel conditions, transmitter is
assumed to send the information at a fixed rate. Under these
assumptions, a two-state (ON-OFF) transmission model is adopted,
where information is transmitted reliably at a fixed rate in the ON
state while no reliable transmission occurs in the OFF state. QoS
limitations are imposed as constraints on buffer violation
probabilities, and effective capacity formulation is used to
identify the maximum throughput that a wireless channel can sustain
while satisfying statistical QoS constraints. Energy efficiency is
investigated by obtaining the bit energy required at zero spectral
efficiency and the wideband slope in both wideband and low-power
regimes assuming that the receiver has perfect channel side
information (CSI). In the wideband regime, it is shown that the bit
energy required at zero spectral efficiency is the minimum bit
energy. A similar result is shown for a certain class of fading
distributions in the low-power regime. In both wideband and
low-power regimes, the increased energy requirements due to the
presence of QoS constraints are quantified. Comparisons with
variable-rate/fixed-power and variable-rate/variable-power cases are
given.

Energy efficiency is further analyzed in the presence of channel
uncertainties. The scenario in which \emph{a priori} unknown fading
coefficients are estimated at the receiver via minimum
mean-square-error (MMSE) estimation with the aid of training
symbols, is considered. The optimal fraction of power allocated to
training is identified under QoS constraints. It is proven that the
minimum bit energy in the low-power regime is attained at a certain
nonzero power level below which bit energy increases without bound
with vanishing power. Hence, it is shown that it is extremely energy
inefficient to operate at very low power levels when the channel is
only imperfectly known.


\end{abstract}
\newpage

\setcounter{page}{1}
\begin{spacing}{1.6}
\section{Introduction}

The two key characteristics of wireless communications that most
greatly impact system design and performance are 1) the
randomly-varying channel conditions and 2) limited energy resources.
In wireless systems, the power of the received signal fluctuates
randomly over time due to mobility, changing environment, and
multipath fading caused by the constructive and destructive
superimposition of the multipath signal components \cite{book}.
These random changes in the received signal strength lead to
variations in the instantaneous data rates that can be supported by
the channel. In addition, mobile wireless systems can only be
equipped with limited energy resources, and hence energy efficient
operation is a crucial requirement in most cases.

To measure and compare the energy efficiencies of different systems
and transmission schemes, one can choose as a metric the energy
required to reliably send one bit of information.
Information-theoretic studies show that energy-per-bit requirement
is generally minimized, and hence the energy efficiency is
maximized, if the system operates at low signal-to-noise ratio
($\tsnr$) levels and hence in the low-power or wideband regimes.
Recently, Verd\'u in \cite{sergio} has determined the minimum bit
energy required for reliable communication over a general class of
channels, and studied of the spectral efficiency--bit energy
tradeoff in the wideband regime while also providing novel tools
that are useful
for analysis at low $\tsnr$s. 

In many wireless communication systems, in addition to
energy-efficient operation, satisfying certain quality of service
(QoS) requirements is of paramount importance in providing
acceptable performance and quality. For instance, in voice over IP
(VoIP), interactive-video (e.g,. videoconferencing), and
streaming-video applications in wireless systems, latency is a key
QoS metric and should not exceed certain levels \cite{Szigeti}. On
the other hand, wireless channels, as described above, are
characterized by random changes in the channel, and such volatile
conditions present significant challenges in providing QoS
guarantees. In most cases, statistical, rather than deterministic,
QoS assurances can be given.


In summary, it is vital for an important class of wireless systems
to operate efficiently while also satisfying QoS requirements (e.g.,
latency, buffer violation probability). Information theory provides
the ultimate performance limits and identifies the most efficient
use of resources. However, information-theoretic studies and Shannon
capacity formulation generally do not address delay and quality of
service (QoS) constraints \cite{Ephremides}. Recently, Wu and Negi
in \cite{dapeng} defined the effective capacity as the maximum
constant arrival rate that a given time-varying service process can
support while providing statistical QoS guarantees. Effective
capacity formulation uses the large deviations theory and
incorporates the statistical QoS constraints by capturing the rate
of decay of the buffer occupancy probability for large queue
lengths. The analysis and application of effective capacity in
various settings has attracted much interest recently (see e.g.,
\cite{wu-downlink}--\cite{deli} and references therein). For
instance,  Tang and Zhang in \cite{tang-powerrate} considered the
effective capacity when both the receiver and transmitter know the
instantaneous channel gains, and derived the optimal power and rate
adaptation technique that maximizes the system throughput under QoS
constraints. These results are extended to multichannel
communication systems in \cite{tang-multichannel}. Liu \emph{et al.}
in \cite{finite} considered fixed-rate transmission schemes and
analyzed the effective capacity and related resource requirements
for Markov wireless channel models. In this work, the
continuous-time Gilbert-Elliott channel with ON and OFF states is
adopted as the channel model while assuming the fading coefficients
as zero-mean Gaussian distributed. A study of cooperative networks
operating under QoS constraints is provided in
\cite{liu-cooperation}. In \cite{deli}, we have investigated the
energy efficiency under QoS constraints by analyzing the normalized
effective capacity (or equivalently the spectral efficiency) in the
low-power and wideband regimes. We considered
variable-rate/variable-power and variable-rate/fixed-power
transmission schemes assuming the availability of channel side
information at both the transmitter and receiver or only at the
receiver.


In this paper, we consider a wireless communication scenario in
which only the receiver has the channel side information, and the
transmitter, not knowing the channel conditions, sends the
information at a fixed-rate with fixed power. If the fixed-rate
transmission cannot be supported by the channel, we assume that
outage occurs and information has to be retransmitted. Similarly as
in \cite{finite}, we consider a channel model with ON and OFF
states. In this scenario, we investigate the energy efficiency under
QoS constraints in the low-power and wideband regimes by considering
the bit energy requirement defined as average energy normalized by
the effective capacity. Our analysis will initially be carried out
under the assumption that the receiver has perfect channel
information. Subsequently, we consider the scenario in which \emph{a
priori} unknown channel is estimated by the receiver with the
assistance of training symbols, albeit only imperfectly.


The rest of the paper is organized as follows. Section II introduces
the system model. In Section III, we briefly describe the notion of
effective capacity and the spectral efficiency--bit energy tradeoff.
Assuming the availability of the perfect channel knowledge at the
receiver, we analyze the energy efficiency in the wideband and
low-power regimes in Sections \ref{sec:wideband} and
\ref{sec:lowpower}, respectively. In Section \ref{sec:imperfect}, we
investigate the energy efficiency in the low-power regime when the
receiver knows the channel only imperfectly. Finally, Section
\ref{sec:conclusion} concludes the paper.

\section{System Model} \label{sec:model}

\begin{figure}
\begin{center}
\includegraphics[width=\figsize\textwidth]{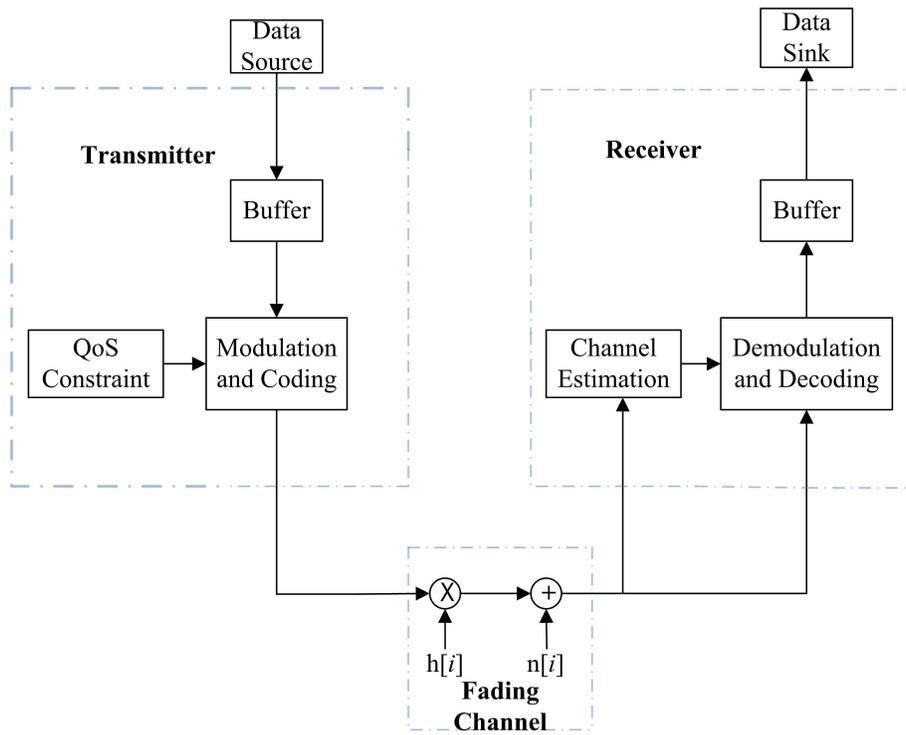}
\caption{The general system model.}\label{fig:0}
\end{center}
\end{figure}

We consider a point-to-point wireless link in which there is one
source and one destination. The system model is depicted in Figure
\ref{fig:0}. It is assumed that the source generates data sequences
which are divided into frames of duration $T$. These data frames are
initially stored in the buffer before they are transmitted over the
wireless channel. The discrete-time channel input-output relation in
the $i^{\text{th}}$ symbol duration is given by
\begin{gather} \label{eq:model}
y[i] = h[i] x[i] + n[i] \quad i = 1,2,\ldots.
\end{gather}
where $x[i]$ and $y[i]$ denote the complex-valued channel input and
output, respectively. We assume that the bandwidth available in the
system is $B$ and the channel input is subject to the following
average energy constraint: $\E\{|x[i]|^2\}\le \Pb / B$ for all $i$.
Since the bandwidth is $B$, symbol rate is assumed to be $B$ complex
symbols per second, indicating that the average power of the system
is constrained by $\Pb$. Above in (\ref{eq:model}), $n[i]$ is a
zero-mean, circularly symmetric, complex Gaussian random variable
with variance $\E\{|n[i]|^2\} = N_0$. The additive Gaussian noise
samples $\{n[i]\}$ are assumed to form an independent and
identically distributed (i.i.d.) sequence. Finally, $h[i]$ denotes
the channel fading coefficient, and $\{h[i]\}$ is a stationary and
ergodic discrete-time process. We denote the magnitude-square of the
fading coefficients by $z[i]=|h[i]|^2$.

In this paper, we initially consider the scenario in which
the receiver has perfect channel side
information and hence perfectly knows the instantaneous values of
$\{h[i]\}$ while the transmitter has no such knowledge. Subsequently, we will analyze the effect of imperfect channel knowledge at the receiver. When the receiver perfectly knows the channel conditions, the instantaneous channel capacity with channel
gain $z[i] \!\!= \!\!|h[i]|^2$ is
\begin{equation}\label{eq:shannon}
C[i]=B\log_2(1+\tsnr z[i]) \text{ bits/s}
\end{equation}
where $\tsnr = \Pb/(N_{0}B)$ is the average transmitted
signal-to-noise ratio. Since the transmitter is unaware of the
channel conditions, information is transmitted at a fixed rate of
$r$ bits/s. When $r < C$, the channel is considered to be in the
ON state and reliable communication is achieved at this rate. If, on
the other hand, $r \ge C$, outage occurs. In this case, channel is in
the OFF state and reliable communication at the rate of $r$ bits/s
cannot be attained. Hence, effective data rate is zero and
information has to be resent. We assume that a simple automatic
repeat request (ARQ) mechanism is incorporated in the communication
protocol to acknowledge the reception of data and to ensure that the
erroneous data is retransmitted \cite{finite}.

\begin{figure}
\begin{center}
\includegraphics[width=\figsize\textwidth]{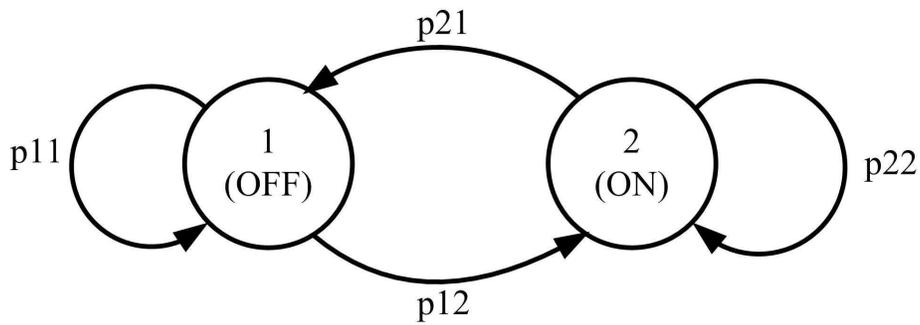}
\caption{ON-OFF state transition model.}\label{fig:00}
\end{center}
\end{figure}

Fig. \ref{fig:00} depicts the two-state transmission model together
with the transition probabilities. In this paper, we assume that the
channel fading coefficients stay constant over the frame duration
$T$. Hence, the state transitions occur at every $T$ seconds. Now,
the probability of staying in the ON state, $p_{22}$, is defined as
follows\footnote{The formulation in (\ref{eq:p_22}) assumes as
before that the symbol rate is $B$ symbols/s and hence we have $TB$
symbols in a duration of $T$ seconds.}:
\begin{align}
p_{22} &= P\{r < C[i+TB] \, \big | \, r < C[i]\} = P\{z[i+TB]
> \alpha \, \big | \, z[i] > \alpha \} \label{eq:p_22}
\end{align}
where
\begin{equation}\label{eq:thresh}
\alpha=\frac{2^{\frac{r}{B}}-1}{\tsnr}.
\end{equation}
Note that $p_{22}$ depends on the joint distribution of $(z[i+TB],
z[i])$. For the Rayleigh fading channel, the joint density function
of the fading amplitudes can be obtained in closed-form
\cite{rapajic}. In this paper, with the goal of simplifying the
analysis and providing results for arbitrary fading distributions,
we assume that fading realizations are independent for each
frame\footnote{This assumption also enables us to compare the
results of this paper with those in \cite{deli} in which
variable-rate/variable-power and variable-rate/fixed-power
transmission schemes are studied for block fading channels.}. Hence,
we basically consider a block-fading channel model. Note that in
block-fading channels, the duration $T$ over which the fading
coefficients stay constant can be varied to model fast or slow
fading scenarios.

Under the block fading assumption, we now have $p_{22} = P\{z[i+TB]
> \alpha\} = P\{z > \alpha\}$. Similarly, the other transition
probabilities become
\begin{align}
p_{11}&=p_{21}=P\{z \le \alpha\}=\int_0^\alpha p_z(z)dz\\
p_{22}&=p_{12}=P\{z >  \alpha\}=\int_\alpha^\infty p_z(z)dz
\end{align}
where $p_z$ is the probability density function of $z$. We finally note that
$rT$ bits are successfully transmitted and received in the ON state,
while the effective transmission rate in the OFF state is zero.

\section{Preliminaries -- Effective Capacity and Spectral Efficiency-Bit Energy Tradeoff}

In \cite{dapeng}, Wu and Negi defined the effective capacity as the
maximum constant arrival rate\footnote{For time-varying arrival
rates, effective capacity specifies the effective bandwidth of the
arrival process that can be supported by the channel.} that a given
service process can support in order to guarantee a statistical QoS
requirement specified by the QoS exponent $\theta$. If we define $Q$
as the stationary queue length, then $\theta$ is the decay rate of
the tail distribution of the queue length $Q$:
\begin{equation}
\lim_{q \to \infty} \frac{\log P(Q \ge q)}{q} = -\theta.
\end{equation}
Therefore, for large $q_{\max}$, we have the following approximation
for the buffer violation probability: $P(Q \ge q_{\max}) \approx
e^{-\theta q_{max}}$. Hence, while larger $\theta$ corresponds to
more strict QoS constraints, smaller $\theta$ implies looser QoS
guarantees. Moreover, if $D$ denotes the steady-state delay
experienced in the buffer, then it is shown in \cite{liu-isit} that $P\{D \ge d_{\max}\} \le c \sqrt{P\{Q\ge q_{\max}\}}$ for constant arrival rates. This result provides a link between the buffer and delay violation probabilities. In the above formulation, $c$ is some positive constant, $q_{\max} = a d_{\max}$, and $a$ is the source arrival rate.


Now, the effective capacity for a given QoS exponent $\theta$ is
obtained from
\begin{gather}
-\lim_{t\rightarrow\infty}\frac{1}{\theta
t}\log_e{\mathbb{E}\{e^{-\theta S[t]}\}} \stackrel{\text{def}}{=}
-\frac{\Lambda(-\theta)}{\theta}
\end{gather}
where $S[t] = \sum_{k=1}^{t}R[k]$ is the time-accumulated service
process and $\{R[k], k=1,2,\ldots\}$ denote the discrete-time,
stationary and ergodic stochastic service process. Note that in the
model we consider, $R[k] = rT \text{ or } 0$ depending on the
channel state being ON or OFF, respectively. In \cite{chang} and
\cite[Section 7.2, Example 7.2.7]{cschang}, it is shown that for such an ON-OFF model, we have
%
\begin{equation}\label{eq:bookresult}
\frac{\Lambda(\theta)}{\theta}=\frac{1}{\theta}\log_e\Big(\frac{1}{2}\Big(p_{11}+p_{22}e^{\theta
Tr}+\sqrt{(p_{11}+p_{22}e^{\theta Tr})^2+4(p_{11}+p_{22}-1)e^{\theta
Tr}} \Big)\Big).
\end{equation}
Using the formulation in (\ref{eq:bookresult}) and noting that $p_{11}+p_{22}=1$ in our  model, we express the effective capacity normalized by the frame
duration $T$ and bandwidth $B$, or equivalently spectral efficiency
in bits/s/Hz, for a given statistical QoS
constraint $\theta$, as
\begin{align}
\hspace{-.2cm}\R_E(\tsnr,
\theta)=\frac{1}{TB}\max_{r\geq0}\Big\{-\frac{\Lambda(-\theta)}{\theta}\Big\}
&=\max_{r\geq0}\Big\{-\frac{1}{\theta
TB}\log_e\big(p_{11}+p_{22}e^{-\theta
Tr}\big)\Big\}\\
&=\max_{r\geq0}\Big\{-\frac{1}{\theta
TB}\log_e\big(1-P\{z>\alpha\}(1-e^{-\theta
Tr})\big)\Big\}\label{eq:specec}\\
&=-\frac{1}{\theta TB}\log_e\Big(1-P\{z>\alphao\}\big(1-e^{-\theta
T\ro}\big)\Big)\,\, \text{bits/s/Hz}\label{eq:spececopt}
\end{align}
where $\ro$ is the maximum fixed transmission rate that solves
(\ref{eq:specec}) and $\alphao = (2^{\frac{\ro}{B}}-1)/\tsnr$. Note
that both $\alphao$ and $\ro$ are functions of $\tsnr$ and $\theta$.

The normalized effective capacity, $\R_E$, provides the maximum
throughput under statistical QoS constraints in the fixed-rate
transmission model. It can be easily shown that
\begin{gather} \label{eq:R_Etheta0}
\lim_{\theta \to 0} \R_E(\tsnr,\theta) = \max_{r\ge 0}
\,\,\frac{r}{B} \, P\{z > \alpha\}.
\end{gather}
Hence, as the QoS requirements relax, the maximum constant arrival
rate approaches the average transmission rate. On the other hand,
for $\theta > 0$, $\R_E < \frac{1}{B} \max_{r\ge 0} r P\{z
>\alpha\}$ in order to avoid violations of QoS constraints.

%
%


In this paper, we focus on the energy efficiency of wireless
transmissions under the aforementioned statistical QoS limitations. Since energy efficient operation generally requires operation   at low-$\tsnr$ levels, our analysis throughout the paper is carried out
in the low-$\tsnr$ regime. In
this regime, the tradeoff between the normalized effective capacity
(i.e, spectral efficiency) $\R_E$ and bit energy $\frac{E_b}{N_0} =
\frac{\tsnr}{\R_E(\tsnr)}$ is a key tradeoff in understanding the
energy efficiency, and is characterized by the bit energy at zero
spectral efficiency and wideband slope provided, respectively, by
\begin{equation}\label{eq:ebresult}
\frac{E_b}{N_0}\bigg|_{\R = 0}= \lim_{\tsnr \to 0} \frac{\tsnr}{\R_E(\tsnr)}= \frac{1}{\dot{\R}_{E}(0)} \text{ and
} \mathcal{S}_0=-\frac{2(\dot{\R_E}(0))^2}{\ddot{\R_E}(0)}\log_e{2}
\end{equation}
where $\dot{\R}_E(0)$ and $\ddot{\R}_E(0)$ are the first and second
derivatives with respect to $\tsnr$, respectively, of the function $\R_E(\tsnr)$ at zero $\tsnr$ \cite{sergio}.
$\frac{E_b}{N_0}\Big|_{\R=0}$ and $\mathcal{S}_0$ provide a linear
approximation of the spectral efficiency curve at low spectral
efficiencies, i.e.,
\begin{gather}\label{eq:linearapprox}
\R_E\left(\frac{E_b}{N_0}\right) = \frac{\mathcal{S}_0}{10\log_{10}
2} \left( \frac{E_b}{N_0}\bigg|_{dB}-\frac{E_b}{N_0}\bigg|_{\R = 0,
dB}\right) + \epsilon
\end{gather}
where $\frac{E_b}{N_0}\Big|_{dB} = 10\log_{10}\frac{E_b}{N_0}$ and
$\epsilon = o\left( \frac{E_b}{N_0}-\frac{E_b}{N_0}\Big|_{\R =
0}\right)$. Moreover, $\frac{E_b}{N_0}\Big|_{\R=0}$ is the minimum
bit energy $\frac{E_b}{N_0}_{\tmin}$ when the spectral efficiency
$\R_E$ is a non-decreasing concave function of $\tsnr$. Indeed, we
show that when the channel is perfectly known at the receiver,
$\frac{E_b}{N_0}\Big|_{\R=0} = \frac{E_b}{N_0}_{\tmin}$ in the
wideband regime as $B \to \infty$. Moreover, we demonstrate that
$\frac{E_b}{N_0}\Big|_{\R=0} = \frac{E_b}{N_0}_{\tmin}$ for Rayleigh
and Nakagami fading channels (with integer fading parameter $m$) in
the low-power regime as $\Pb \to 0$.
On the other hand, for general treatment, we refer to the bit energy
required as $\tsnr$ vanishes as $\frac{E_b}{N_0}\Big|_{\R=0}$
throughout the paper. As we shall see in Section \ref{sec:imperfect}
that $\frac{E_b}{N_0}\Big|_{\R=0}$ is not necessarily the minimum
bit energy in a certain scenario of the imperfectly-known channel.

\section{Energy Efficiency in the Wideband Regime} \label{sec:wideband}

In this section, we consider the wideband regime in which the
bandwidth is large. We assume that the average power $\Pb$ is kept
constant. Note that as the bandwidth $B$ increases, $\tsnr =
\frac{\Pb}{N_0B}$ approaches zero and we operate in the low-$\tsnr$
regime.

We first introduce the notation $\zeta=\frac{1}{B}$. Note
that as $B \to \infty$, we have $\zeta \to 0$. Moreover, with this
notation, the normalized effective capacity can be expressed as\footnote{Since the results in the paper are generally obtained for fixed but arbitrary $\theta$, the normalized effective capacity is often expressed in the paper as $\R_E(\tsnr)$ instead of $\R_E(\tsnr,\theta)$ to avoid cumbersome expressions.}
\begin{equation}\label{eq:spececb}
\hspace{-0.03cm}\R_E(\tsnr)=-\frac{\zeta}{\theta
T}\log_e\Big(1-P\{z>\alphao\}\big(1-e^{-\theta T\ro}\big)\Big).
\end{equation}
Note that $\alphao$ and $\ro$ are also in general dependent on $B$
and hence $\zeta$. The following result provides the expressions for
the bit energy at zero spectral efficiency (i.e., as $B \to \infty$)
and the wideband slope, and characterize the spectral efficiency-bit
energy tradeoff in the wideband regime.
\begin{Lem} \label{theo:wideband}
In the wideband regime, the bit energy at zero spectral efficiency,
and wideband slope are given by
\begin{gather}
\frac{E_b}{N_0}\bigg|_{\R = 0}=\frac{-\delta\log_e2}{\log_e\xi} \quad \text{and} \label{eq:ebminwb}\\
\mathcal{S}_0=\frac{2\xi\log_e^2\xi}{(\delta\sta)^2
P\{z>\sta\}e^{-\delta\sta}},\label{eq:s0wb}
\end{gather}
respectively, where $\delta=\frac{\theta T\Pb}{N_0\log_e2}$ and
$\xi=1-P\{z>\sta\}(1-e^{-\delta\sta})$. $\sta$ is defined
as
$\sta=\lim_{\zeta\rightarrow0}\alphao$
and $\sta$ satisfies
\begin{equation} \label{eq:stacondwideband}
\delta\sta=\log_e\left(1+\delta\frac{P\{z>\sta\}}{p_z(\sta)}\right).
\end{equation}
\end{Lem}
\emph{Proof: } Assume that the Taylor series expansion of $\ro$ with
respect to small $\zeta$ is
\begin{equation}\label{eq:substRb}
\ro=\stro+\dot{r}_{\text{opt}}(0)\zeta+o(\zeta)
\end{equation}
where $\stro=\lim_{\zeta\rightarrow0}\ro$ and
$\dot{r}_{\text{opt}}(0)$ is the first derivative with respect to
$\zeta$ of $\ro$  evaluated at $\zeta=0$. From (\ref{eq:thresh}), we
can find that
\begin{align}
\alphao=\frac{2^{\ro\zeta}-1}{\frac{\Pb\zeta}{N_0}}=\frac{\stro
\log_e2}{\frac{\Pb}{N_0}}+\frac{\dot{r}_{\text{opt}}(0)\log_e2+\frac{(\stro
\log_e2)^2}{2}}{\frac{\Pb}{N_0}}\,\zeta+o(\zeta)
\end{align}
from which we have as $\zeta\rightarrow0$ that
\begin{equation}\label{eq:stab}
\sta=\frac{\stro \log_e2}{\frac{\Pb}{N_0}}
\end{equation}
and that
\begin{equation}\label{eq:dotalpha}
\dao(0)=\frac{\dro(0)\log_e2+\frac{(\stro
\log_e2)^2}{2}}{\frac{\Pb}{N_0}}
\end{equation}
where $\dao(0)$ is the first derivative with respect to $\zeta$ of
$\alphao$ evaluated at $\zeta=0$. According to (\ref{eq:stab}),
$\stro=\frac{\Pb\sta}{N_0\log_e2}$. We now have
\begin{align}
\frac{E_b}{N_0}\bigg|_{\R =
0}&=\lim_{\zeta\rightarrow0}\frac{\frac{\Pb}{N_0} \zeta}{\R_E(\zeta)} = \frac{\frac{\Pb}{N_0}}{\dot{\R}_E(0)}=\frac{-\frac{\theta
T\Pb}{N_0}}{\log_e\big(1-P\{z>\sta\}(1-e^{-\theta
T\stro})\big)}
=\frac{-\delta\log_e2}{\log_e\xi}\label{eq:ebminb}
\end{align}
where $\dot{\R}_E(0)$ is the derivative of $\R_E$ with respect to $\zeta$ at $\zeta = 0$, $\delta=\frac{\theta T\Pb}{N_0\log_e2}$, and
$\xi=1-P\{z>\sta\}(1-e^{-\delta\sta})$. Therefore, we prove
(\ref{eq:ebminwb}).
%
Note that the second derivative $\ddot{\R}_E(0)$, required in the
computation of the wideband slope $\mathcal{S}_0$, can be obtained
from
\begin{align}
\ddot{\R}_E(0)&=\lim_{\zeta\rightarrow0}2\frac{\R_E(\zeta)-\dot{\R}_E(0)\zeta}{\zeta^2}\nonumber\\
&=\lim_{\zeta\rightarrow0}2\frac{1}{\zeta}\Big(-\frac{1}{\theta
T}\log_e\big(1-P\{z>\alphao\}\big(1-e^{-\theta T\ro}\big)\big)
+\frac{1}{\theta T}\log_e\big(1-P\{z>\sta\}(1-e^{-\theta
T\stro})\big)\Big)\nonumber\\
&=\lim_{\zeta\rightarrow0}-\frac{2}{\theta
T}\frac{\big(p_z(\alphao)\dao(\zeta)(1-e^{-\theta
T\ro})
-P\{z>\alphao\}\theta Te^{-\theta
T\ro}\dro(\zeta)\big)}{1-P\{z>\alphao\}\big(1-e^{-\theta
T\ro}\big)}
\label{eq:ddotproof1}\\
&=-\frac{2}{\theta T}\frac{\big(p_z(\sta)\dao(0)(1-e^{-\theta
T\stro})
-P\{z>\sta\}\theta Te^{-\theta
T\stro}\dro(0)\big)}{1-P\{z>\sta\}\big(1-e^{-\theta T\stro
}\big)}
\label{eq:ddotproof2}
\end{align}
where $\stro=\frac{\Pb\sta}{N_0\log_e2}$. Above,
(\ref{eq:ddotproof1}) and (\ref{eq:ddotproof2}) follow by using
L'Hospital's Rule and applying Leibniz Integral Rule \cite{Protter}.%


Next, we derive an equality satisfied by $\sta$. Consider the
objective function in (\ref{eq:specec})
\begin{gather}\label{eq:objfunc}
-\frac{1}{\theta TB}\log_e\big(1-P\{z>\alpha\}(1-e^{-\theta
Tr})\big).
\end{gather}
It can easily be seen that both as $r \to 0$ and $r \to \infty$,
this objective function approaches zero\footnote{Note that $\alpha$
increases without bound with increasing $r$.}. Hence,
(\ref{eq:objfunc}) is maximized at a finite and nonzero value of $r$
at which the derivative of (\ref{eq:objfunc}) with respect to $r$ is
zero. 
Differentiating (\ref{eq:objfunc})
with respect to $r$ and making it equal to zero leads to the
equality that needs to be satisfied at the optimal value $\ro$:
\begin{equation}\label{eq:conditionb}
\frac{2^{\ro\zeta}p_z(\alphao)N_0\log_e2}{\Pb}(1-e^{-\theta
T\ro})=\theta Te^{-\theta T\ro}P\{z>\alphao\}
\end{equation}
where $\zeta = 1/B$.
For given $\theta$, as the bandwidth increases (i.e.,
$\zeta\rightarrow0$), $\ro \to \stro$. Clearly, $\stro \neq 0$ in
the wideband regime.
 Because, otherwise, if $\ro \to 0$ and consequently $\alphao\rightarrow0$, the left-hand-side of
(\ref{eq:conditionb}) becomes zero, while the right-hand-side is
different from zero. So, employing (\ref{eq:stab}) and taking the
limit of both sides of (\ref{eq:conditionb}) as $\zeta \to 0$, we
can derive that
\begin{equation}\label{eq:conditionbrev}
\frac{p_z(\sta)N_0\log_e2}{\Pb}\left(1-e^{-\frac{\theta
T\Pb}{N_0\log_e2}\sta}\right)=\theta Te^{-\frac{\theta
T\Pb}{N_0\log_e2}\sta}P\{z>\sta\}
\end{equation}
which, after rearranging, yields
\begin{equation}\label{eq:subsalphab}
\frac{\theta T\Pb}{N_0\log_e2}\sta=\log_e\left(1+\frac{\theta
T\Pb}{N_0\log_e2}\frac{P\{z>\sta\}}{p_z(\sta)}\right).
\end{equation}
Denoting $\delta=\frac{\theta T\Pb}{N_0\log_e2}$, we obtain the
condition (\ref{eq:stacondwideband}) stated in the theorem.

Combining (\ref{eq:conditionbrev}) and (\ref{eq:dotalpha}) with
(\ref{eq:ddotproof2}) gives us
\begin{align}
\ddot{\R}_E(0)&=-\frac{N_0\log_e^22}{\theta T\Pb}\frac{\stro^2
p_z(\sta)(1-e^{-\theta T\stro})}{1-P\{z>\sta\}\big(1-e^{-\theta
T\stro }\big)}
=-\frac{\stro^2P\{z>\sta\} e^{-\theta
T\stro}\log_e2}{1-P\{z>\sta\}\big(1-e^{-\theta T\stro
}\big)}\label{ddotec}
\end{align}
Substituting (\ref{ddotec}) and the expression for $\dot{\R}_E(0)$ in (\ref{eq:ebminb}) into
(\ref{eq:ebresult}), we obtain (\ref{eq:s0wb}). \hfill$\square$

The following result shows that in the wideband regime,
$\frac{E_b}{N_0}\Big|_{\R = 0}$ is the indeed the minimum bit
energy.

\begin{Lem}\label{theo:minenergywideband}
In the wideband regime, the bit energy required at zero spectral
efficiency (i.e., bit energy required as $B \to \infty$ or
equivalently as $\zeta \to 0$) is the minimum bit energy, i.e.,
\begin{gather}
\frac{E_b}{N_0}\bigg|_{\R = 0}=\frac{-\delta\log_e2}{\log_e\xi} =
\frac{E_b}{N_0}_{\min}.
\end{gather}
\end{Lem}

\emph{Proof:} Since $\frac{E_b}{N_0} =
\frac{\frac{\Pb}{N_0}}{\frac{\R_E(\zeta)}{\zeta}}$, we can show the
result by proving that $\R_E(\zeta)/\zeta$ monotonically decreases
with increasing $\zeta$, and hence achieves its maximum as $\zeta \to 0$. We first evaluate the first derivative of
$\R_E(\zeta)/\zeta$ with respect to $\zeta$:
\begin{align}
\frac{d(\R_E(\zeta)/\zeta)}{d\zeta} &=-\frac{1}{\theta
T}\frac{\frac{p_z(\alphao)N_0}{\Pb}\frac{2^{\ro\zeta}(\dro\zeta+\ro)\zeta\log_e2-(2^{\ro\zeta}-1)}{\zeta^2}(1-e^{-\theta
T\ro})-\theta Te^{-\theta T\ro}\dro
P(z>\alphao)}{1-P(z>\alphao)(1-e^{-\theta
T\ro})}\\
&=-\frac{p_z(\alphao)N_0}{\theta
T\Pb}\frac{2^{\ro\zeta}\ro\zeta\log_e2-(2^{\ro\zeta}-1)}{\zeta^2}
\label{eq:R_Ederiv}
\end{align}
where (\ref{eq:R_Ederiv}) is obtained by using the equation
$\frac{2^{\ro/B}p_z(\alphao)\log_e2}{B\tsnr}(1-e^{-\theta
T\ro})=\theta Te^{-\theta T\ro}P\{z>\alphao\}$ that needs to be
satisfied by $\ro$ and $\alphao$ as shown in the proof of Theorem
\ref{theo:wideband} in (\ref{eq:conditionb}). Note that the
probability density function $p_z(z) \ge 0$ for all $z \ge 0$.
Hence, if $2^{\ro\zeta}\ro\zeta\log_e2-(2^{\ro\zeta}-1) \ge 0$ for
all $\ro \ge 0$ and $\zeta \ge 0$, then
$\frac{d(\R_E(\zeta)/\zeta)}{d\zeta}\le 0$ proving that
$\R_E(\zeta)/\zeta$ is indeed a monotonically decreasing function of
$\zeta$. Now, we denote $x=\ro\zeta\geq0$ and define
$f(x)=2^x x\log_e2-(2^x-1)$. The first derivative of $f$ with respect
to $x$ is $\dot{f}(x)=x2^x(\log_e2)^2\geq0$, implying that $f$ is
a monotonically increasing function. Since $f(0) = 0$, we
immediately conclude that $f(x) \ge 0$ for all $x \ge 0$. Hence,
$2^{\ro\zeta}\ro\zeta\log_e2-(2^{\ro\zeta}-1) \ge 0$ for all $\ro
\ge 0$ and $\zeta \ge 0$,
$\frac{d(\R_E(\zeta)/\zeta)}{d\zeta}\leq0$, and
$\frac{E_b}{N_0}_{\tmin}$ is achieved in the limit as $\zeta \to 0$.
\hfill $\square$

\begin{figure}
\begin{center}
\includegraphics[width=\figsize\textwidth]{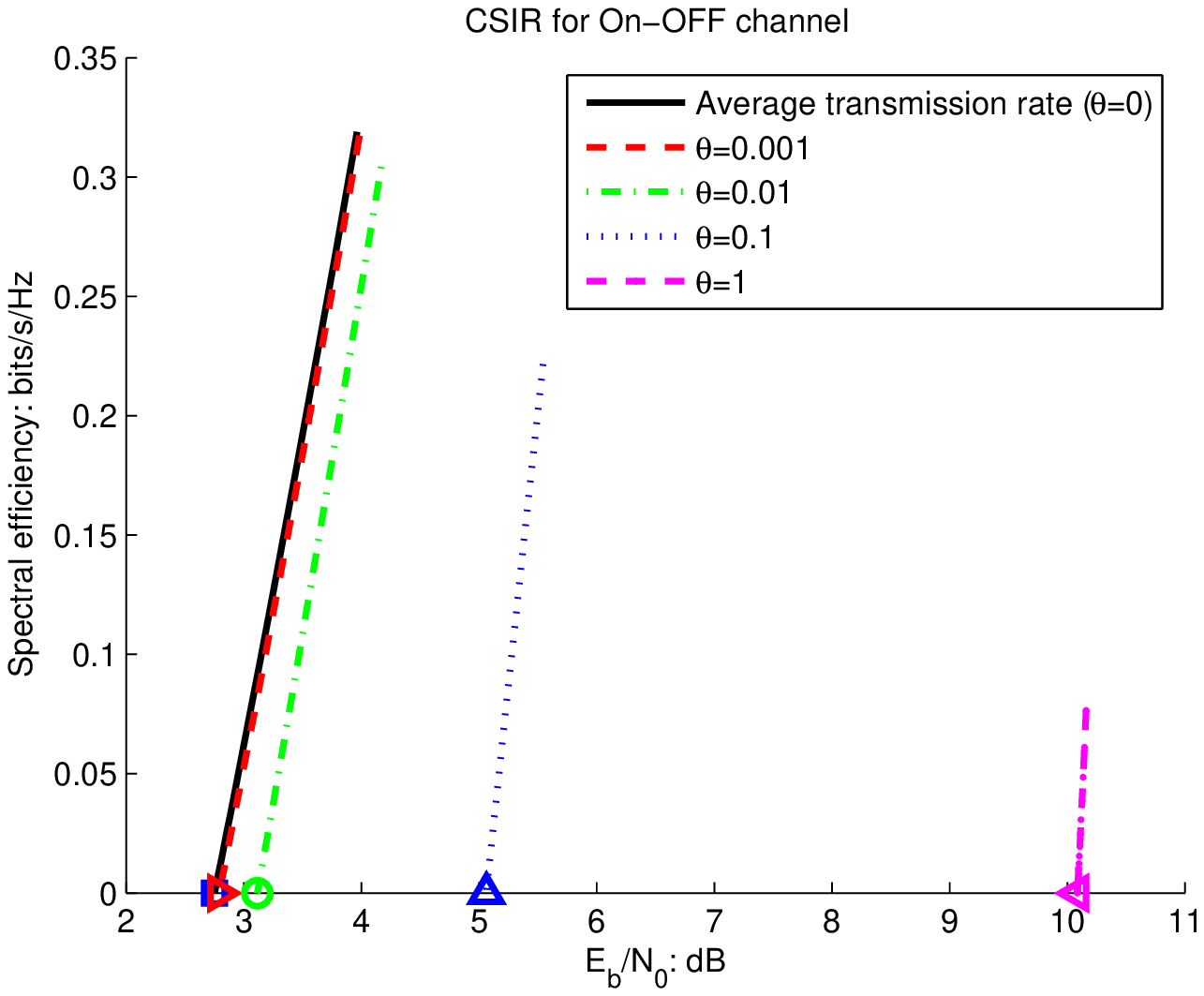}
\caption{Spectral efficiency vs. $E_b/N_0$ in the Rayleigh
channel.}\label{fig:4}
\end{center}
\end{figure}

\begin{figure}
\begin{center}
\includegraphics[width=\figsize\textwidth]{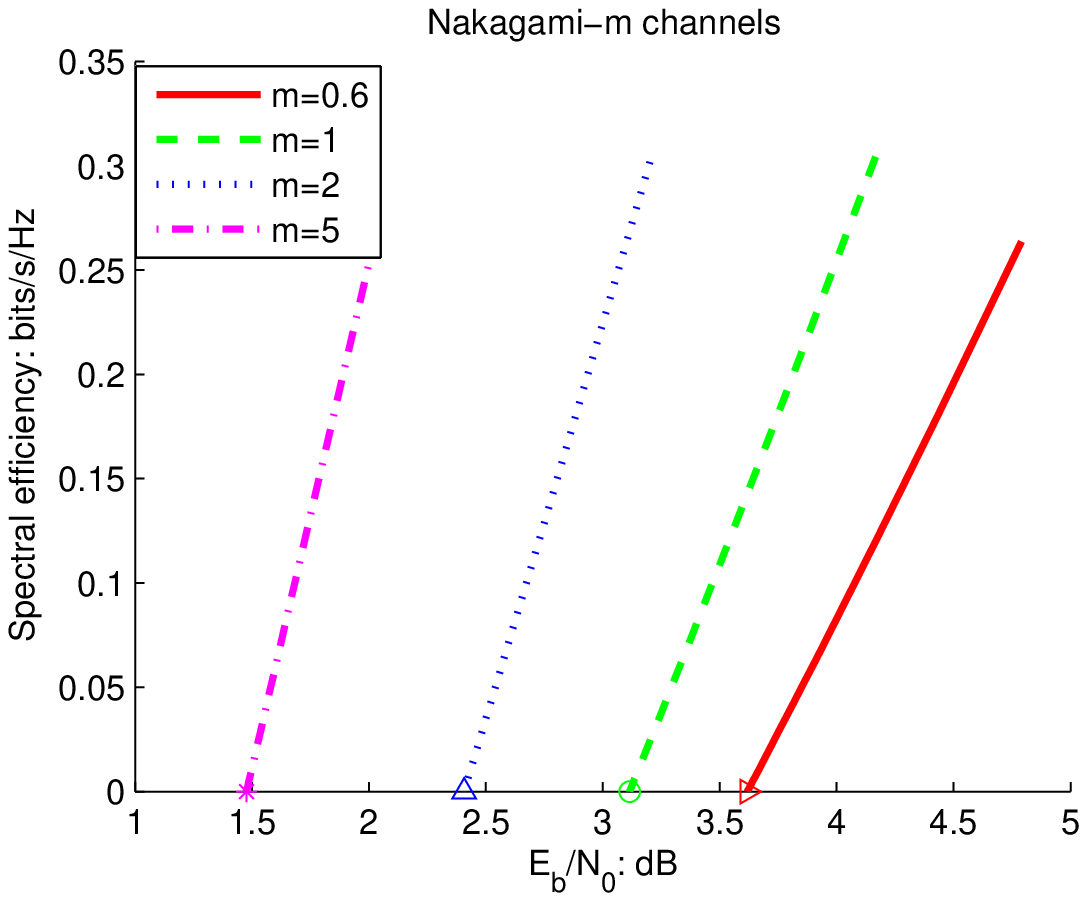}
\caption{Spectral efficiency vs. $E_b/N_0$ in Nakagami-$m$ channels;
$\theta=0.01$, $m = 0.6, 1, 2, 5$.}\label{fig:wbm}
\end{center}
\end{figure}

Having analytically characterized the spectral efficiency--bit
energy tradeoff in the wideband regime, we now provide numerical
results to illustrate the theoretical findings. Fig. \ref{fig:4}
plots the spectral efficiency curves as a function of the bit energy
in the Rayleigh channel. In all the curves, we have
$\Pb/N_{0}=10^4$. Moreover, we set $T = 2$ ms in the numerical
results throughout the paper. As predicted by the result of Theorem
\ref{theo:minenergywideband}, $\frac{E_b}{N_0}\Big|_{\R = 0} =
\frac{E_b}{N_0}_{\tmin}$ in all cases in Fig. \ref{fig:4}. It can be
found
that $\sta=\{1, 0.9858,0.8786,0.4704,0.1177\}$ from which we obtain $\frac{E_b}{N_0}_{\tmin} = \{2.75, 2.79,    3.114,     5.061,   10.087\}$dB for $\theta = \{0, 0.001,0.01,0.1,1\}$, respectively. 
For the same set of $\theta$ values in the same sequence, we compute
the wideband slope values as $\mathcal{S}_0=\{0.7358,
0.7463,0.8345,1.4073,3.1509\}$. We immediately observe that more
stringent QoS constraints and hence higher values of $\theta$ lead
to higher minimum bit energy values and also higher energy
requirements at other nonzero spectral efficiencies. Fig.
\ref{fig:wbm} provides the spectral efficiency curves for
Nakagami-$m$ fading channels for different values of $m$. In this
figure, we set $\theta = 0.01$. For $m = 0.6, 1, 2, 5$, we find that
$\sta = \{1.0567,  0.8786    0.7476    0.6974\}$,
$\frac{E_b}{N_0}_{\tmin} =\{3.618,    3.114,    2.407,    1.477\}$,
and $\mathcal{S}_0 = \{0.6382,    0.8345,    1.1220,    1.4583\}$,
respectively. Note that as $m$ increases and hence the channel
conditions improve, the minimum bit energy decreases and the
wideband slope increases, improving the energy efficiency both at
zero spectral efficiency and at nonzero but small spectral
efficiency values. As $m \to \infty$, the performance approaches
that of the unfaded additive Gaussian noise channel (AWGN) for which
we have $\frac{E_b}{N_0}_{\tmin} = -1.59$ dB and $\mathcal{S}_0 = 2$
\cite{sergio}.

\section{Energy Efficiency in the Low-Power Regime} \label{sec:lowpower}

In this section, we investigate the spectral efficiency--bit energy
tradeoff as the average power $\Pb$ diminishes. We assume that the
bandwidth allocated to the channel is fixed. Note that $\tsnr =
\Pb/(N_0B)$ vanishes with decreasing $\Pb$, and we again operate in
the low-$\tsnr$ regime similarly as in Section \ref{sec:wideband}.
However, energy requirements in the low-power regime will be
different from those in the wideband regime, because the arrival rates that can be
supported get smaller with decreasing power in this regime.

The following result provides the expressions for the bit energy at zero spectral
efficiency and the wideband slope.

\begin{Lem} \label{theo:lowpower}
In the low-power regime, the bit energy at zero spectral efficiency
and wideband slope are given by \vspace{-.5cm}
\begin{gather}
\frac{E_b}{N_0}\bigg|_{\R = 0}=\frac{\log_e2}{\sta P\{z>\sta\}} \quad \text{and} \label{eq:ebminlp}\\
\mathcal{S}_0=\frac{2 P\{z>\sta\}}{1+\beta
(1-P\{z>\sta\})},\label{eq:s0lp}
\end{gather}
respectively, where $\beta=\frac{\theta TB}{\log_e2}$ is normalized
QoS constraint. In the above formulation, $\sta$ is again defined as $
\sta = \lim_{\tsnr \to 0} \alphao, $ and $\sta$ satisfies
\begin{gather} \label{eq:stacond}
\sta p_z(\sta)=P\{z>\sta\}.
\end{gather}
\end{Lem}
\indent \emph{Proof:} We first consider the Taylor series expansion
of $\ro$ in the low-$\tsnr$ regime:
\begin{equation}\label{eq:subsR}
\ro=a\tsnr+b\tsnr^2+o(\tsnr^2)
\end{equation}
where $a$ and $b$ are real-valued constants. Substituting
(\ref{eq:subsR}) into (\ref{eq:thresh}), we obtain the Taylor series
expansion for $\alphao$:
\begin{equation}\label{eq:subsalpha}
\alphao =
\frac{a\log_e2}{B}+\left(\frac{b\log_e2}{B}+\frac{a^2\log_e^2
2}{2B^2}\right)\tsnr+o(\tsnr).
\end{equation}
>From (\ref{eq:subsalpha}), we note that in the limit as $\tsnr \to
0$, we have
\begin{equation}\label{eq:coeffRa}
\sta=\frac{a\log_e2}{B}.
\end{equation}
Next, we obtain the Taylor series expansion with respect to $\tsnr$
for $P\{z>\alphao\}$ using the Leibniz Integral Rule \cite{Protter}:
\begin{align}\label{eq:probalpha}
P\{z>\alphao\}
=P\{z>\sta\}-\left(\frac{b\log_e2}{B}+\frac{a^2\log_e^2
2}{2B^2}\right)
p_z(\sta)\tsnr+o(\tsnr).
\end{align}
Using (\ref{eq:subsR}), (\ref{eq:subsalpha}), and
(\ref{eq:probalpha}),  we find the following series expansion for
$\R_E$ given in (\ref{eq:spececopt}):
\begin{align}\label{eq:reducespec}
\hspace{-0.2cm}\R_E(\tsnr)&=-\frac{1}{\theta
TB}\log_e\Bigg[1-\bigg(P\{z>
\sta\}-\bigg(\frac{b\log_e2}{B}+\frac{a^2\log_e^2
2}{2B^2}\bigg)p_z(\sta)\tsnr+o(\tsnr)\bigg)\nonumber\\
&\phantom{-\frac{1}{\theta T}-1-\bigg(P\{z>\sta\}}\times\big(\theta
Ta\tsnr+(\theta
Tb-\frac{(\theta Ta)^2}{2})\tsnr^2+o(\tsnr^2)\big)\Bigg]\nonumber\\
&=\frac{a P\{z>\sta\}}{B}\tsnr+\frac{1}{B}\Big(-\frac{\theta
Ta^2}{2}P\{z>\sta\}-\frac{a^3p_z(\sta)\log_e^2
2}{2B^2}+\frac{\theta T(P\{z>\sta\}a)^2}{2}\Big)\tsnr^2
+o(\tsnr^2). 
\end{align}
Then, using (\ref{eq:coeffRa}), we immediately derive from
(\ref{eq:reducespec}) that 
\begin{align}
\dot{\R}_E(0)&=\frac{\sta P\{z>\sta\}}{\log_e2},\label{eq:dotec}\\
\ddot{\R}_E(0)&=
-\frac{\sta^3 p_z\{\sta\}}{\log_e2}-\frac{\theta TB\sta^2}{\log_e^2
2}P\{z>\sta\}(1-P\{z>\sta\}).\label{eq:ddotec}
\end{align}
Similarly as in the discussion in the proof of Theorem \ref{theo:wideband} in Section
\ref{sec:wideband}, the optimal fixed-rate $\ro$, akin to
(\ref{eq:conditionb}), should  satisfy
\begin{equation}\label{eq:constthr}
\frac{2^{\ro/B}p_z(\alphao)\log_e2}{B\tsnr}(1-e^{-\theta
T\ro})=\theta Te^{-\theta T\ro}P\{z>\alphao\}.
\end{equation}
Taking the limits of both sides of (\ref{eq:constthr}) as $\tsnr \to
0$ and employing (\ref{eq:subsR}), we obtain
\begin{equation}\label{eq:condition}
\frac{a p_z(\sta)\log_e2}{B}=P\{z>\sta\}.
\end{equation}
>From (\ref{eq:coeffRa}), (\ref{eq:condition}) simplifies to
\begin{equation}\label{eq:reducedalpha}
\sta p_z(\sta)=P\{z>\sta\},
\end{equation}
proving the condition in (\ref{eq:stacond}). Moreover,
using (\ref{eq:reducedalpha}), the first term in the expression for
$\ddot{\R}_E(0)$ in (\ref{eq:ddotec}) becomes $-\frac{\sta^2
P\{z\geq\sta\}}{\log_e2}$. Together with this change, evaluating the
expressions in (\ref{eq:ebresult}) with the results in
(\ref{eq:dotec}) and (\ref{eq:ddotec}), we obtain (\ref{eq:ebminlp})
and (\ref{eq:s0lp}). \hfill $\square$

Next, we show that the equation (\ref{eq:stacond}) that needs to be
satisfied by $\sta$ has a unique solution for a certain class of
fading distributions.

\begin{Lem} \label{theo:stacond}
The equation $ \sta p_z(\sta)=P\{z>\sta\} $ has a unique solution
when $z$ has a Gamma distribution with integer parameter $n$, i.e.,
when the probability density function of $z$ is given by
\begin{gather}\label{eq:gammadensity0}
p_z(z) = \frac{\lambda^n}{\Gamma(n)} z^{n-1} e^{-\lambda z}
\end{gather}
where $n \ge 1$ is an integer, $\lambda > 0$, and $\Gamma$ is the
Gamma function \cite{Grimmett}.
\end{Lem}

\emph{Proof}: See Appendix \ref{app:stacond}.

\emph{Remark:} In the special case in which $n = \lambda = m$ and $m
\ge 1$ is an integer, the Gamma density (\ref{eq:gammadensity0})
becomes
\begin{equation}
p_z(z)=\frac{m^m z^{m-1}}{\Gamma(m)}e^{-mz}
\end{equation}
which is the probability density function of $z = |h|^2$ in
Nakagami-$m$ fading channels (with integer $m$)\cite{book}.
Moreover, when $m = 1$, we have the Rayleigh fading channel in which
$z$ has an exponential distribution, i.e., $p_{z}(z) = e^{-z}$.
Therefore, the result of Theorem \ref{theo:stacond} applies for
these channels.

\emph{Remark:} Theorem \ref{theo:lowpower} shows that the
$\frac{E_b}{N_0}\Big|_{\R = 0}$ for any $\theta \ge 0$ depends only
on $\sta$. From Theorem \ref{theo:stacond}, we know that if $z$ has
the Gamma density function given by (\ref{eq:gammadensity0}), then
$\sta$ is unique and hence is the same for all $\theta \ge 0$. We
immediately conclude from these results that
$\frac{E_b}{N_0}\Big|_{\R = 0}$ also has the same value for all
$\theta \ge 0$ and therefore does not depend on $\theta$ when $z$
has the distribution given in (\ref{eq:gammadensity0}).

Moreover, using the results of Theorem  \ref{theo:stacond} above and
Theorem \ref{theo:minenergywideband} in Section \ref{sec:wideband},
we can further show that $\frac{E_b}{N_0}\Big|_{\R = 0}$ is the
minimum bit energy.  Note that this implies that the same minimum
bit energy can be attained regardless of how strict the QoS
constraint is. On the other hand, we note that the wideband slope
$\mathcal{S}_0$ in general varies with $\theta$.

\begin{Corr}
In the low-power regime, when $\theta = 0$, the minimum bit energy
is achieved as $\Pb \to 0$, i.e., $\frac{E_b}{N_0}\Big|_{\R = 0} =
\frac{E_b}{N_0}_{\min}$. Moreover, if the probability density
function of $z$ is in the form given in (\ref{eq:gammadensity0})
then the minimum bit
energy is achieved as $\Pb \to 0$, i.e. $\frac{E_b}{N_0}\Big|_{\R =
0} = \frac{E_b}{N_0}_{\min}$, for all $\theta \ge 0$.
\end{Corr}

\emph{Proof:} Recall from (\ref{eq:R_Etheta0}) that in the limit as
$\theta \to 0$,
\begin{gather}
\R_E(\tsnr,0) = \lim_{\theta \to 0} \R_E(\tsnr,\theta) = \max_{r\ge
0} \,\,\frac{r}{B} \, P\left\{z >
\frac{2^{\frac{r}{B}}-1}{\tsnr}\right\}.
\end{gather}
Since the optimization is performed over all $r \ge 0$, it can be
easily seen that the above maximization problem can be recast as
follows:
\begin{gather}\label{eq:R_Etheta0equiv}
\R_E(\tsnr,0) = \max_{x\ge 0} \,\,x \, P\left\{z >
\frac{2^x-1}{\tsnr}\right\}.
\end{gather}
>From (\ref{eq:R_Etheta0equiv}), we note that $\R_E(\tsnr,0)$ depends
on $B$ only through $\tsnr = \frac{\Pb}{N_0 B}$. Therefore,
increasing $B$ has the same effect as decreasing $\Pb$. Hence,
low-power and wideband regimes are equivalent when $\theta = 0$.
Consequently, the result of Theorem \ref{theo:minenergywideband},
which shows that the minimum bit energy is achieved as $B \to
\infty$, implies that the minimum bit energy is also achieved as
$\Pb \to 0$.

Note that $\R_E(\tsnr,\theta) \le \R_E(\tsnr,0)$ for $\theta > 0$.
Therefore, the bit energy required when $\theta > 0$ is larger than
that required when $\theta = 0$. On the other hand, as we have
proven in Theorem \ref{theo:stacond}, $\sta$ is unique and the bit
energy required as $\Pb \to 0$ is the same for all $\theta \ge 0$
when $z$ has a Gamma density in the form given in
(\ref{eq:gammadensity0}). Since the minimum bit energy in the case
of $\theta = 0$ is achieved as $\Pb \to 0$, and the same bit energy
is attained for all $\theta > 0$, we immediately conclude that
$\frac{E_b}{N_0}\Big|_{\R = 0} = \frac{E_b}{N_0}_{\min}$ for all
$\theta \ge 0$ when $z$ has a Gamma distribution. \hfill $\square$

\begin{figure}
\begin{center}
\includegraphics[width=\figsize\textwidth]{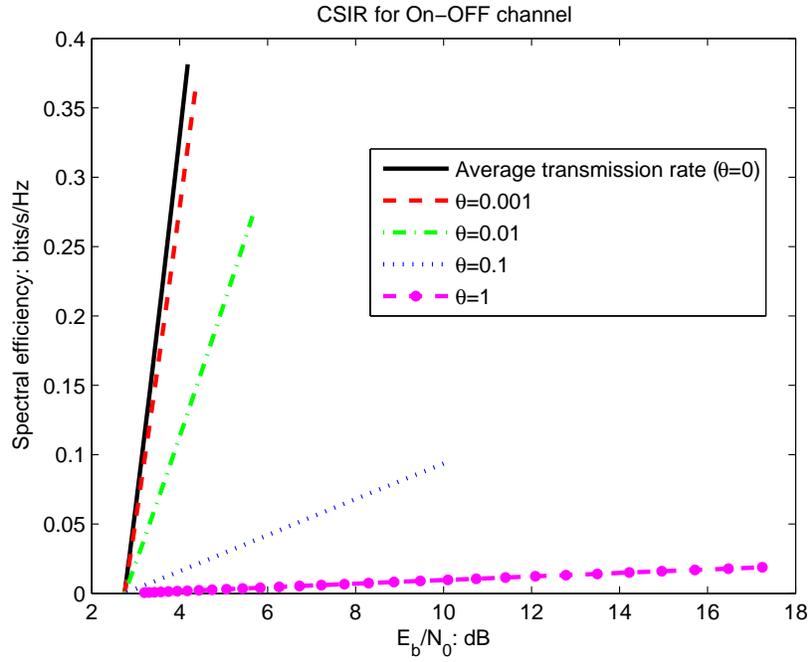}
\caption{Spectral efficiency vs. $E_b/N_0$ in the Rayleigh channel
(equivalently Nakagami-$m$ channel with $m=1$).}\label{fig:2}
\end{center}
\end{figure}

Next, we provide numerical results which confirm the theoretical
conclusions and illustrate the impact of QoS constraints on the
energy efficiency. We set $B=10^5$
Hz in the computations. Fig. \ref{fig:2} plots the spectral efficiency as a function of
the bit energy for different values of $\theta$ in the Rayleigh
fading channel (or equivalently Nakagami-$m$ fading channel with $m
= 1$) for which $\E\{|h|^2\}=\E\{z\}=1$.
In all cases in Fig. \ref{fig:2}, we readily note that
$\frac{E_b}{N_0}\Big|_{\R = 0} = \frac{E_b}{N_0}_{\tmin}$. Moreover,
as predicted, the minimum bit energy is the same and is equal to the
one achieved when there are no QoS constraints (i.e., when $\theta =
0$). From the equation $\sta p_z(\sta)=P\{z>\sta\}$, we can find
that $\sta=1$ in the Rayleigh channel for which $p_z(\sta) = P\{z
> \sta\} = e^{-\sta}$. Hence, the minimum bit energy is
$\frac{E_b}{N_0}_{\min} = 2.75$ dB. On the other hand, the wideband
slopes are $\mathcal{S}_0=\{0.7358, 0.6223,0.2605,0.0382,0.0040\}$
for $\theta=\{0, 0.001,0.01,0.1,1\}$, respectively. Hence,
$\mathcal{S}_0$ decreases with increasing $\theta$ and consequently
more bit energy is required at a fixed nonzero spectral efficiency.
Assuming that the minimum bit energies are the same and considering
the linear approximation in (\ref{eq:linearapprox}), we can easily
show for fixed spectral efficiency $\R\left(\frac{E_b}{N_0}\right)$
for which the linear approximation is accurate that the increase in
the bit energy in dB, when the QoS exponent increases from
$\theta_1$ to $\theta_2$, is
\begin{gather}
\frac{E_b}{N_0}\bigg|_{dB,
\theta_2}-\frac{E_b}{N_0}\bigg|_{dB,\theta_1} =
\left(\frac{1}{\mathcal{S}_{0,\theta_2}} -
\frac{1}{\mathcal{S}_{0,\theta_1}}\right)
\R\left(\frac{E_b}{N_0}\right) 10\log_{10}2.
\label{eq:bitenergydifference}
\end{gather}
As observed in Fig. \ref{fig:2} (and also as will be seen in Fig.
\ref{fig:3} below), spectral efficiency curves are almost linear in
the low-power regime, validating the accuracy of the linear
approximation in (\ref{eq:linearapprox}) obtained through
$\frac{E_b}{N_0}\Big|_{\R = 0}$ and $\mathcal{S}_0$.

\begin{figure}
\begin{center}
\includegraphics[width=\figsize\textwidth]{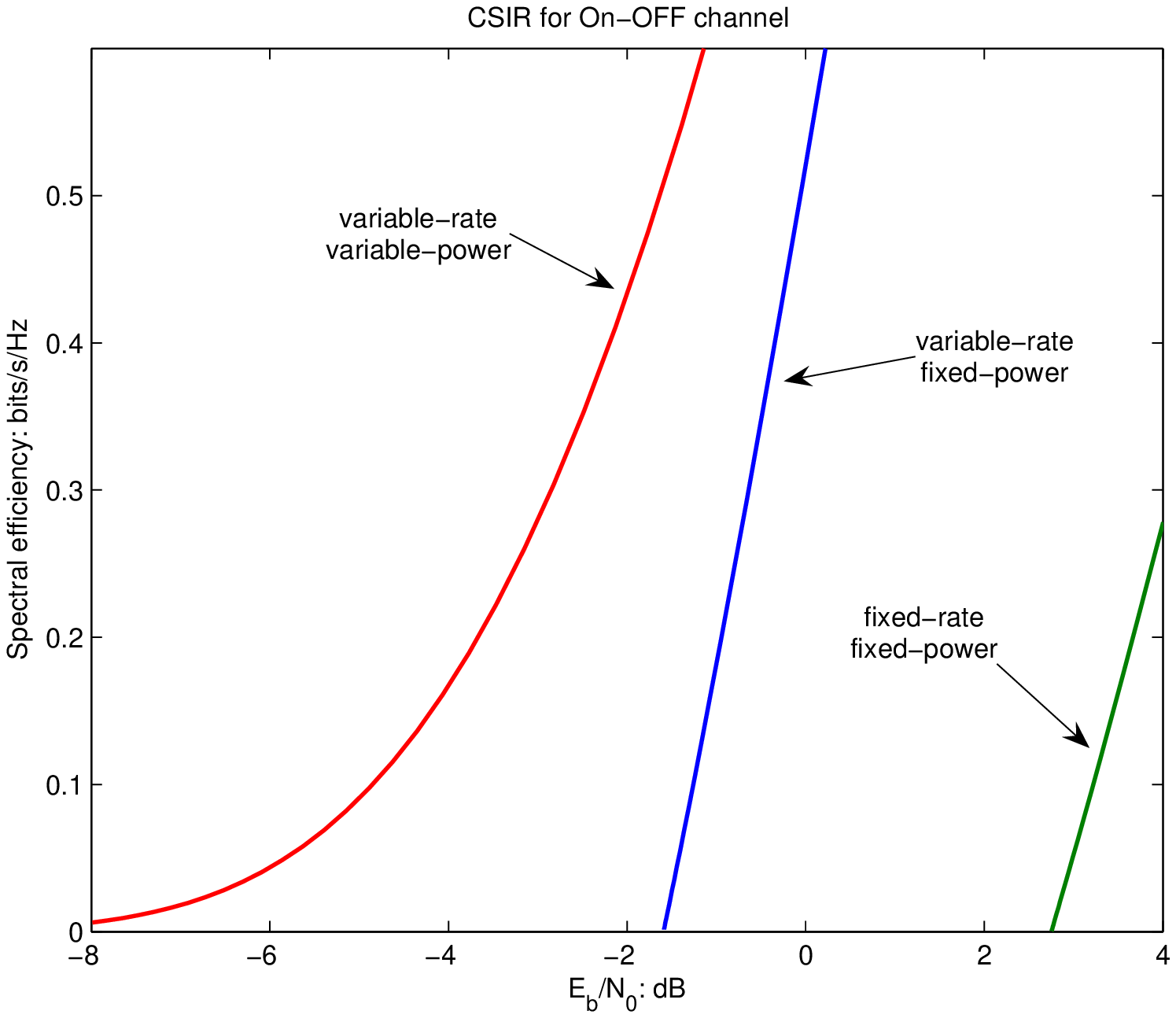}
\caption{Spectral efficiency vs. $E_b/N_0$ in the Rayleigh channel;
$\theta=0.001$.}\label{fig:spec-comp}
\end{center}
\end{figure}

Fig. \ref{fig:3} plots the spectral efficiency curves as a function
of the bit energy for Nakagami-$m$ channels for different values of $m$.  $\theta$ is set to be 0.01. For $m=\{0.6,1,2,5\}$, we compute that
$\sta=\{1.2764,1,0.809,0.7279\}$,  $\frac{E_b}{N_0}_{\min} = \{3.099,    2.751,    2.176,    1.343\}$, and $\mathcal{S}_0=\{0.1707,0.2605,0.4349,0.7479\}$, respectively. We
observe that as $m$ increases and hence the channel quality
improves, lower bit energies are required. Finally, in Fig.
\ref{fig:spec-comp}, we plot the spectral efficiency vs. $E_b/N_0$
for different transmission strategies. The
variable-rate/variable-power and variable-rate/fixed-power
strategies are studied in \cite{deli}. We immediately see that
substantially more energy is required for fixed-rate/fixed-power
transmission schemes considered in this paper.

\section{The Effect of Imperfect Channel Knowledge in the Low-Power Regime}
\label{sec:imperfect}

In this section, as a major difference from the previous sections,
we consider the scenario in which neither the transmitter nor the
receiver has channel side information prior to transmission.
Moreover, we consider a particular fading distribution and assume
that the fading coefficients are zero-mean Gaussian random variables
with variance $E\{|h|^2\} = E\{z\} = \gamma$. We further assume that
the system operates in two phases: training phase and data
transmission phase. In the training phase, known pilot symbols are
transmitted to enable the receiver to estimate the channel
conditions, albeit imperfectly. Following the training phase, data
is transmitted, and the receiver, equipped with the estimate of the
channel, attempts to recover the data from the received signal.
Through this scenario, we investigate the effect of the imperfect
channel knowledge on the energy efficiency when the system is
subject to QoS constraints. We note that training-based transmission
schemes have received much interest due to their practical
significance (see e.g., \cite{Tong} -- \cite{gursoy} and references
therein).

Under the block-fading assumption, channel estimation has to be
performed every $T$ seconds. We assume that minimum
mean-square-error (MMSE) estimation is employed at the receiver.
Since the MMSE estimate depends only on the training energy and not
on the training duration, it can be easily seen that transmission of
a single pilot at every $T$ seconds is optimal. Note that in every
frame duration of $T$ seconds, we have $TB$ symbols and the overall
available energy is $\Pb T$. We now assume that each frame consists of a pilot symbol and $TB - 1$ data symbols. The energies of the pilot
and data symbols are
\begin{equation}\label{eq:trainpower}
\e_t=\rho \Pb T, \quad\text{and}\quad \e_s=\frac{(1-\rho)\Pb
T}{TB-1},
\end{equation}
respectively, where $\rho$ is the fraction of total energy allocated
to training. Note that the data symbol energy $\e_s$ is obtained by
uniformly allocating the remaining energy among the data symbols.

In the training phase, the receiver obtains the MMSE estimate
$\hat{h}$ which is a circularly symmetric, complex, Gaussian random
variable with mean zero and variance $\frac{\gamma^2 \e_t}{\gamma
\e_t + N_0}$, i.e.,
$\hat{h} \sim \mathcal{CN} \left( 0, \frac{\gamma^2 \e_t}{\gamma
\e_t + N_0} \right)$\cite{gursoy}.
 Now, the
channel fading coefficient $h$ can be expressed as
$h=\hat{h}+\tilde{h}$
where $\tilde{h}$ is the estimate error and $\tilde{h}\sim\mathcal
{CN}(0,\frac{\gamma N_0}{\gamma \e_t+N_0})$. Consequently, the
channel input-output relation becomes
\begin{gather} \label{eq:impmodel}
y[i] = \hat{h}[i] x[i] + \tilde{h}[i] x[i] + n[i] \quad i =
1,2,\ldots.
\end{gather}
Since finding the capacity of the channel in (\ref{eq:impmodel}) is
a difficult task\footnote{In \cite{gursoy}, the capacity of
training-based transmissions under input peak power constraints is
shown to be achieved by an $\tsnr$-dependent, discrete distribution
with a finite number of mass points. In such cases, no closed-form
expression for the capacity exists, and capacity values need to be
obtained through numerical computations.}, a capacity lower bound is
generally obtained by considering the estimate error $\tilde{h}$ as
another source of Gaussian noise and treating $\tilde{h}[i] x[i] +
n[i]$ as Gaussian distributed noise uncorrelated from the input.
Now, the new noise variance is $\E\{|\tilde{h}[i] x[i] + n[i]|^2\} =
\sigma_{\tilde{h}}^2 \e_s + N_0$ where $\sigma_{\tilde{h}}^2 =
\E\{|\tilde{h}|^2\} = \frac{\gamma N_0}{\gamma \e_t+N_0}$ is the
variance of the estimate error. Under these assumptions, a lower
bound on the instantaneous capacity is given by \cite{training},
\cite{gursoy}
\begin{align}
C_L&=\frac{TB-1}{T}\log_2\left(1+ \frac{\e_s}{\sigma_{\tilde{h}}^2
\e_s + N_0} |\hat{h}|^2\right)
=\frac{TB-1}{T} \log_2\left(1+\tsnref |w|^2\right) \text{ bits/s}
\label{eq:traincap2}
\end{align}
where effective $\tsnr$ is
\begin{equation}\label{eq:trainsnr}
\tsnref=\frac{\e_s \sigma_{\hat{h}}^2}{\sigma_{\tilde{h}}^2 \e_s +
N_0},
\end{equation}
and $\sigma^2_{\hat{h}} = \E\{|\hat{h}|^2\} = \frac{\gamma^2
\e_t}{\gamma \e_t + N_0}$ is the variance of estimate $\hat{h}$.
Note that the rightmost expression in (\ref{eq:traincap2}) is obtained
by defining $\hat{h} = \sigma_{\hat{h}} w$ where $w$ is a standard
Gaussian random variable with zero mean and unit variance, i.e., $w\sim\mathcal {CN}(0,1)$.

Since Gaussian is the worst uncorrelated noise \cite{training}, the
above-mentioned assumptions lead to a pessimistic model and   the
rate expression in (\ref{eq:traincap2}) is a lower bound to the
capacity of the true channel (\ref{eq:impmodel}). On the other hand,
$C_L$ is a good measure of the rates achieved in communication
systems that operate as if the channel estimate were perfect (i.e.,
in systems where Gaussian codebooks designed for known channels are
used, and scaled nearest neighbor decoding is employed at the
receiver) \cite{lapidoth}.

Henceforth, we base our analysis on $C_L$ to understand the impact
of the imperfect channel estimate.  Similarly, as in Section
\ref{sec:model}, we assume that the transmitter sends information at
the fixed rate of $r$ bits/s, and the channel is in the ON state if
$r < C_L$. Otherwise, it is in the OFF state. The transition
probabilities are given by \vspace{-.5cm}
\begin{align}
p_{11}=p_{21}=P\{|w|^2 \le \alpha\} \quad \text{and} \quad
p_{22}=p_{12}=P\{|w|^2 > \alpha\}
\end{align}
where
\begin{equation}\label{eq:trainthresh}
\alpha=\frac{2^{\frac{rT}{TB-1}}-1}{\tsnref},
\end{equation}
and $|w|^2$ is an exponential random variable with mean $1$, and hence, $P\{|w|^2 > \alpha\} = e^{-\alpha}$.
Now, the normalized effective capacity is given by
\begin{align}
\R_E(\tsnr,\theta)&=\max_{\substack{r\geq0 \\ 0\leq
\rho\leq1}}{-\frac{1}{\theta
TB}\log_e\big(1-P(|w|^2>\alpha)(1-e^{-\theta
Tr})\big)} \text{ bits/s/Hz} \label{eq:trainopti}\\
&=-\frac{1}{\theta TB}\log_e\big(1-P(|w|^2>\alphao)(1-e^{-\theta
T\ro})\big) \text{ bits/s/Hz} \label{eq:trainopti2}.
\end{align}
Note that $\R_E$ is obtained by optimizing both the fixed
transmission rate $r$ and the fraction of power allocated to
training, $\rho$. In the optimization result (\ref{eq:trainopti2}),
$\ro$ and $\alphao$ are the optimal values of $r$ and $\alpha$,
respectively. We first obtain the following result on the optimal
value of $\rho$.
\begin{Lem} \label{theo:optrho}
At a given $\tsnr$ level, the optimal fraction of power $\rhoo$ that
solves (\ref{eq:trainopti}) does not depend on the QoS exponent
$\theta$ and the transmission rate $r$, and is given by
\begin{equation}\label{eq:optrho}
\rhoo=\sqrt{\eta(\eta+1)}-\eta
\end{equation}
where $\eta=\frac{\gamma TB\tsnr+TB-1}{\gamma TB(TB-2)\tsnr}$ and
$\tsnr = \frac{\Pb}{N_0B}$.
\end{Lem}
\emph{Proof:} From (\ref{eq:trainopti}) and the definition of
$\alpha$ in (\ref{eq:trainthresh}), we can easily see that for fixed
$r$, the only term in (\ref{eq:trainopti}) that depends on $\rho$ is
$\alpha$. Moreover, $\alpha$ has this dependency through $\tsnref$.
Therefore, $\rhoo$ that maximizes the objective function in
(\ref{eq:trainopti}) can be found by minimizing $\alpha$, or
equivalently maximizing $\tsnref$. Substituting the definitions in
(\ref{eq:trainpower}) and the expressions for $\sigma_{\hat{h}}^2$
and $\sigma_{\tilde{h}}^2$ into (\ref{eq:trainsnr}), we have
\begin{equation}\label{eq:trainsnref}
\tsnref=\frac{\e_s \sigma_{\hat{h}}^2}{\sigma_{\tilde{h}}^2 \e_s +
N_0} = \frac{\rho(1-\rho)\gamma^2T^2B^2\tsnr^2}{\rho \gamma
TB(TB-2)\tsnr+\gamma TB\tsnr+TB-1}
\end{equation}
where $\tsnr=\frac{\Pb}{N_0 B}$. Evaluating the derivative of
$\tsnref$ with respect to $\rho$ and making it equal to zero leads to
the expression in (\ref{eq:optrho}). Clearly, $\rhoo$ is independent
of $\theta$ and $r$.

Above, we have implicitly assumed that the maximization is performed
with respect to first $\rho$ and then $r$. However, the result will
not alter if the order of the maximization is changed. Note that the
objective function in (\ref{eq:trainopti})
\begin{gather}
g(\tsnref,r)= - \frac{1}{\theta
TB}\log_e\left(1-P\left(|w|^2>\frac{2^{\frac{rT}{TB-1}}-1}{\tsnref}\right)(1-e^{-\theta
Tr})\right)
\end{gather}
is a monotonically increasing function of $\tsnref$ for all $r$. It
can be easily verified that maximization does not affect the
monotonicity of $g$, and hence $\max_{r \ge 0} g(\tsnref,r)$ is
still a monotonically increasing function of $\tsnref$. Therefore,
in the outer maximization with respect to $\rho$, the choice of
$\rho$ that maximizes $\tsnref$ will also maximize $\max_{r \ge 0}
g(\tsnref,r)$, and the optimal value of $\rho$ is again given by
(\ref{eq:optrho}). \hfill$\square$

\begin{figure}
\begin{center}
\includegraphics[width=\figsize\textwidth]{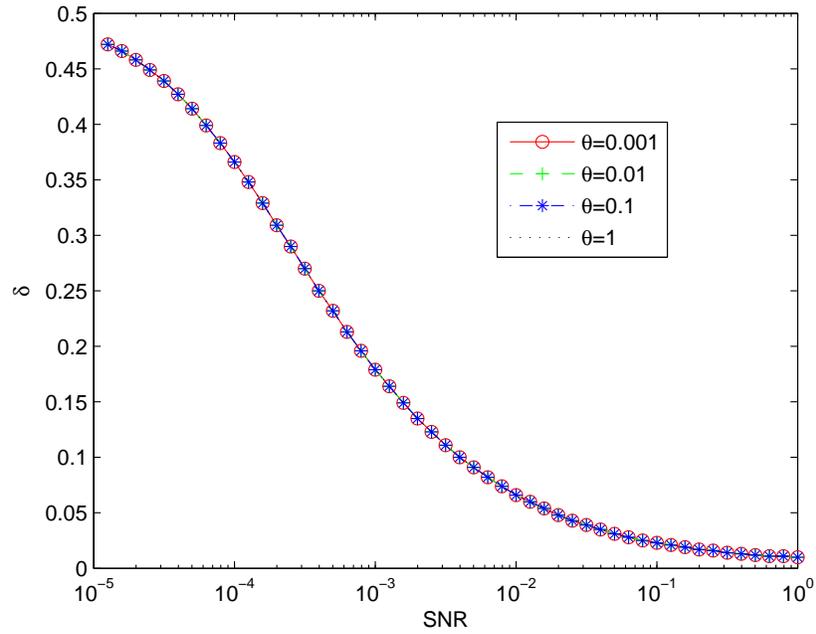}
\caption{Optimal portion $\rhoo$ vs. $\tsnr$ in the Rayleigh
channel. $B=10^7$ Hz.}\label{fig:5}
\end{center}
\end{figure}

Fig. \ref{fig:5} plots $\rhoo$, the optimal fraction of power
allocated to training, as a function of $\tsnr$ for different values
of $\theta$ when $B=10^7$ Hz. As predicted, $\rhoo$ is the same for
all $\theta$. Note that as $\tsnr \to 0$, we have $\eta \to \infty$
and $\rhoo \to 1/2$, which is also observed in the figure. We
further observe in Fig. \ref{fig:5} that $\rhoo$ decreases with
increasing $\tsnr$. Moreover, as $\tsnr \to \infty$, we can find
that $\eta \to \frac{1}{TB-2}$ and hence $\rhoo \to
\sqrt{\frac{1}{TB-2} \left( \frac{1}{TB-2} + 1\right)} -
\frac{1}{TB-2}$. In the figure, we assume $T = 2$ms, and therefore
$TB = 2\times 10^4$ and $\rhoo \to 0.007$.


With the optimal value of $\rho$ given in Theorem \ref{theo:optrho},
we can now express the normalized effective capacity as
\begin{gather}\label{eq:Reimperf}
\R_E(\tsnr,\theta)=\max_{r\geq0 }- \frac{1}{\theta
TB}\log_e\left(1-P\left(|w|^2>\frac{2^{\frac{rT}{TB-1}}-1}{\tsnrefo}\right)(1-e^{-\theta
Tr})\right)
\end{gather}
where \begin{equation}\label{eq:trainsnrefrev}
\tsnrefo=\frac{\phi(\tsnr)\tsnr^2}{\psi(\tsnr)\tsnr+TB-1},
\end{equation}
and
\begin{equation}
\phi(\tsnr)=\rhoo(1-\rhoo)\gamma^2T^2B^2,\text{ and }
\psi(\tsnr)=(1+(TB-2)\rhoo)\gamma TB.
\end{equation}
The formulation in (\ref{eq:Reimperf}) is very similar to that in
(\ref{eq:specec}). The difference is that we have $\alpha =
\frac{2^{\frac{rT}{TB-1}}-1}{\tsnrefo}$ in (\ref{eq:Reimperf}) as
opposed to having $\alpha = \frac{2^{\frac{r}{B}}-1}{\tsnr}$ in
(\ref{eq:specec}). Hence, apart from the change in the scalar that
multiplies $r$ in the expression of $\alpha$, the main difference is
that $\R_E$ in (\ref{eq:Reimperf}) is essentially a function of
$\tsnrefo$. Using this similarity, we obtain the following result
that shows us that operation at very low power levels is extremely
energy inefficient and should be avoided.

\begin{Lem} \label{theo:imperfect}
In the case of imperfectly-known channel, the bit energy increases
without bound as the average power $\Pb$ and hence $\tsnr$ vanishes,
i.e.,
\begin{gather}
\frac{E_b}{N_0}\bigg|_{\R = 0} = \lim_{\tsnr \to 0} \frac{E_b}{N_0}
= \lim_{\tsnr \to 0} \frac{\tsnr}{\R_E(\tsnr)} =
\frac{1}{\dot{\R_E}(0)} = \infty.
\end{gather}
\end{Lem}

\emph{Proof}: As discussed above, the normalized effective capacity
expressions in (\ref{eq:Reimperf}) and (\ref{eq:specec}) are
similar. Essentially, $\R_E$ in (\ref{eq:Reimperf}) can be seen to
be a function of $\tsnrefo$. Then, using the techniques in the proof
of Theorem \ref{theo:lowpower}, we can easily obtain the following
first-order low-$\tsnr$ expansion similar to that in
(\ref{eq:reducespec}):
\begin{align}
\R_E(\tsnr) &= \frac{a P\{|w|^2 > \alphao^*\}}{B} \tsnrefo +
o(\tsnrefo)
\\
&= \frac{TB-1}{TB} \,\,\frac{\alphao^* P\{|w|^2 >
\alphao^*\}}{\log_e2} \tsnrefo + o(\tsnrefo) \label{eq:Reexpansion}
\end{align}
where $\alphao^* = \lim_{\tsnr \to 0} \alphao$. Note that as $\tsnr
\to 0$, $\eta \to \infty$, $\rhoo \to 1/2$, and hence $\phi(\tsnr)
\to 1/4 \gamma^2 T^2 B^2$. Then, we have
\begin{gather}
\tsnrefo = \frac{\gamma^2 T^2 B^2}{4(TB-1)} \tsnr^2 + o(\tsnr^2).
\end{gather}
Hence, $\tsnrefo$ scales as $\tsnr^2$ as $\tsnr$ diminishes to zero,
which implies from (\ref{eq:Reexpansion}) that $\R_E$ scales as
$\tsnr^2$ as well. Therefore, the first derivative of $\R_E$ with
respect to $\tsnr$  is zero at $\tsnr = 0$, i.e., $\dot{\R}_E(0) =
0$, leading to the result that $\lim_{\tsnr \to 0} \frac{E_b}{N_0} =
\infty$. \hfill $\square$

\begin{figure}
\begin{center}
\includegraphics[width=\figsize\textwidth]{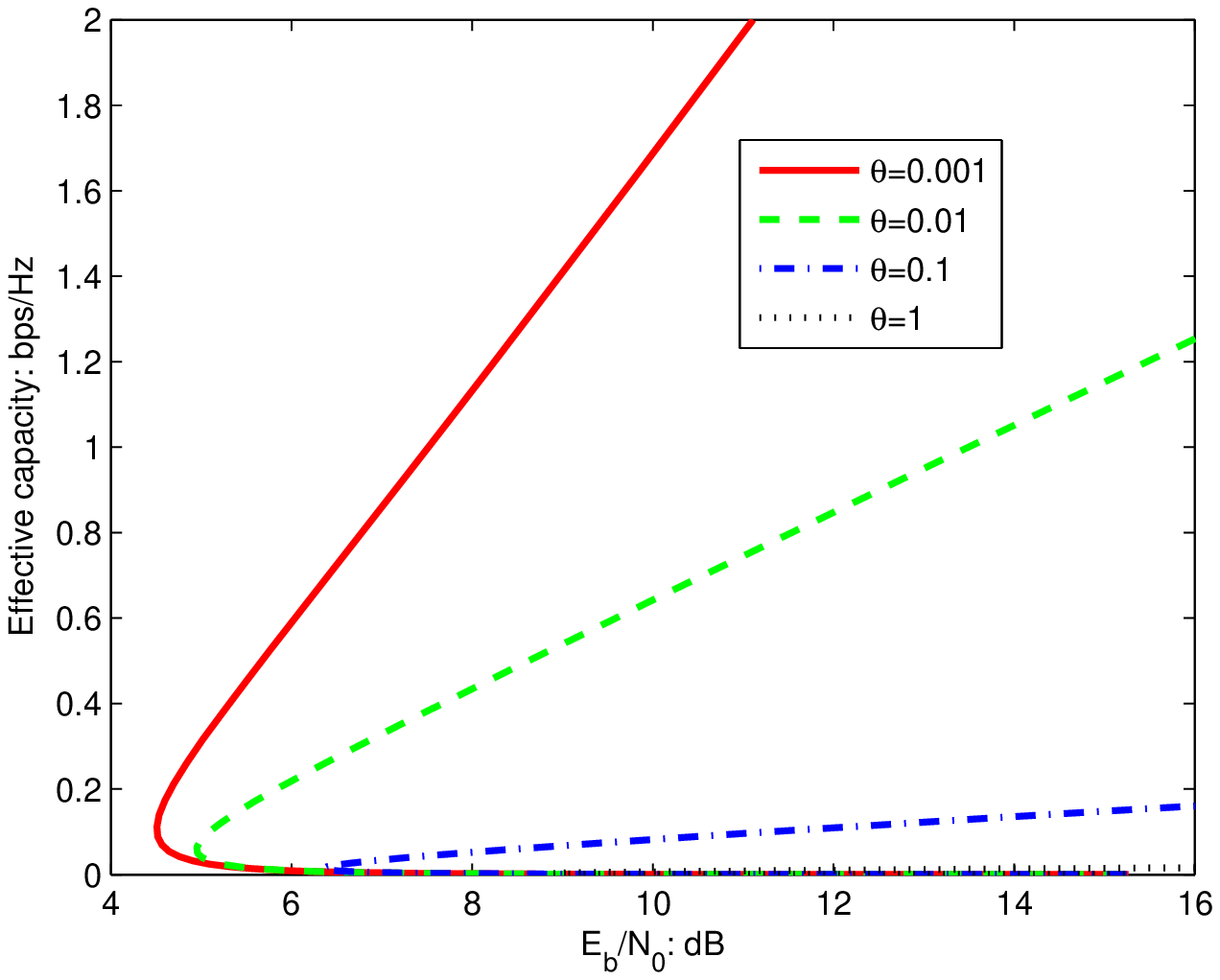}
\caption{Spectral efficiency vs. $E_b/N_0$ in the Rayleigh channel.
$B=10^5$.}\label{fig:6}
\end{center}
\end{figure}

\begin{figure}
\begin{center}
\includegraphics[width=\figsize\textwidth]{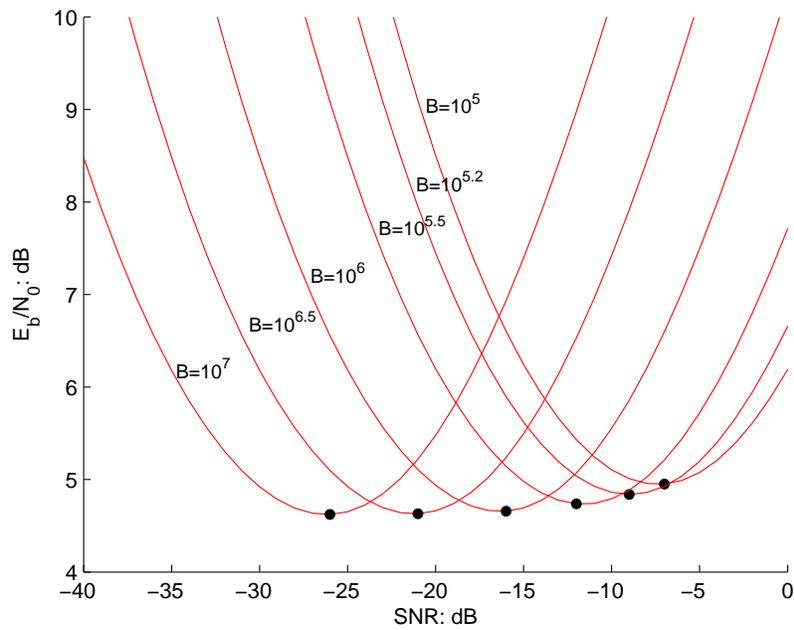}
\caption{$E_b/N_0$ vs. $\tsnr$ in the Rayleigh channel.
$\theta$=0.01.}\label{fig:ebsnr}
\end{center}
\end{figure}

\begin{figure}
\begin{center}
\includegraphics[width=\figsize\textwidth]{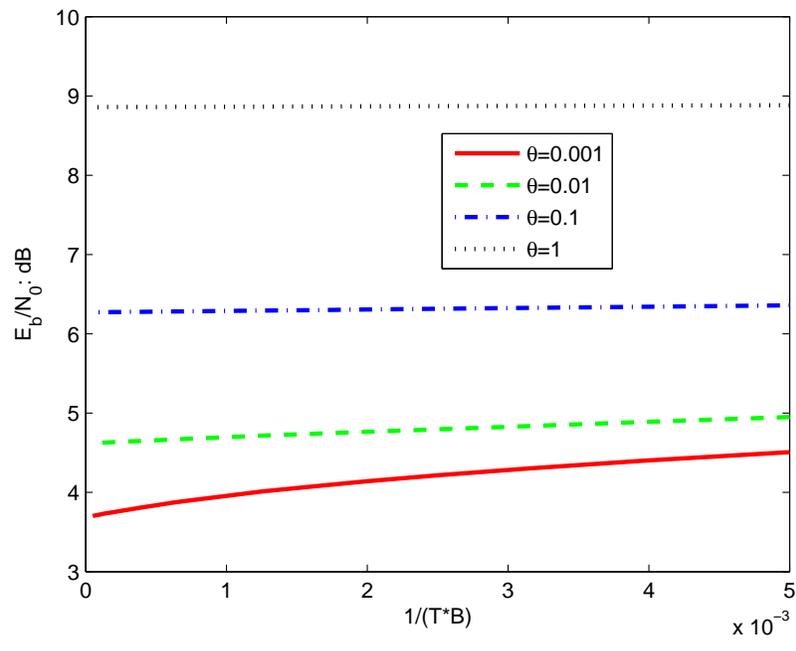}
\caption{$\frac{E_b}{N_0}_{\tmin} $vs. $B$ in the
Rayleigh channel.}\label{fig:7}
\end{center}
\end{figure}

Next, we illustrate the analytical results through numerical computations. Fig.
\ref{fig:6} plots the spectral efficiency vs. bit energy for
$\theta=\{1,0.1,0.01,0.001\}$ when $B=10^5$ Hz. We immediately notice a different behavior as $\tsnr \to 0$ and hence the spectral efficiency decreases. As predicted by the result of Theorem \ref{theo:imperfect}, the bit energy increases without bound as $\R_E \to 0$ in all cases. The minimum bit energy is achieved at a nonzero spectral efficiency below which one should avoid operating as it only increases the energy requirements.
%
In Fig. \ref{fig:ebsnr}, we plot
$\frac{E_b}{N_0}$ as a function of  $\tsnr$ for different bandwidth levels assuming $\theta = 0.01$.
We again observe that the minimum bit energy is attained at a
nonzero $\tsnr$ value below which $\frac{E_b}{N_0}$ requirements
start increasing. Furthermore, we see that as the bandwidth
increases, the minimum bit energy tends to decrease and is achieved
at a lower $\tsnr$ level.
Finally, we plot in Fig. \ref{fig:7} the minimum bit energy as a
function of the bandwidth, $B$. We note that increasing $B$
generally decreases $\frac{E_b}{N_0}_{\min}$ when $\theta = 0$.
However, for the cases in which $\theta > 0$ and there exist QoS
constraints, there is no improvement as $B$ is increased above a
certain value.
\vspace{-0.5cm}
\section{Conclusion} \label{sec:conclusion}

In this paper, we have
considered the effective capacity as a measure of the maximum
throughput under statistical QoS constraints, and analyzed the energy efficiency of fixed-rate transmission schemes over fading channels. In particular, we
have investigated the spectral efficiency--bit energy tradeoff in
the low-power and wideband regimes. We
have obtained expressions for the bit energy at zero spectral efficiency and the wideband
slope, which provide a linear approximation to the spectral efficiency curve at low $\tsnr$s. In the wideband regime, we have shown that the bit energy required at zero spectral efficiency (or equivalently at infinite bandwidth) is the minimum bit energy. We have proven a similar result in the low-power regime for a certain class of fading distributions.   Through this analysis, we
have quantified the increased energy requirements in the presence of
QoS constraints in both wideband and low-power regimes. In the wideband regime, we have noted that the minimum bit energy and wideband slope in general depend on the QoS exponent $\theta$. As the QoS constraints become more stringent and hence $\theta$ is increased, we have observed in the numerical results that the required minimum bit energy increases. On the other hand, in the low power regime, we have shown for a class of fading distributions that the same minimum bit energy is achieved for all $\theta$. However, we have seen that the wideband slope decreases as $\theta$ increases, increasing the energy requirements at nonzero spectral efficiency values.

We have also analyzed energy efficiency in a scenario in which the
fading coefficients are not known prior to transmission and are
estimated imperfectly by the receiver with the aid of training
symbols. We have identified the optimal fraction of power allocated
to training and shown that this optimal fraction do not depend on
the QoS exponent $\theta$ and the transmission rate. In this
scenario, we have further shown that the bit energy requirements
grow without bound in the low-power regime as $\tsnr$ vanishes. This
result shows that the minimum bit energy is attained at a certain
$\tsnr$ value, operating below which should be avoided.


\end{spacing}
\vspace{-0.1cm}
\begin{spacing}{1.1}

\end{spacing}
\newpage

\appendix

\subsection{Proof of Theorem \ref{theo:stacond}} \label{app:stacond}

We consider the following class of Gamma density functions:
\begin{gather} \label{appeq:gammadensity}
p_z(z) = \frac{\lambda^n}{\Gamma(n)} z^{n-1} e^{-\lambda z}
\end{gather}
where $n \ge 1$ is an integer, $\lambda > 0$ and $\Gamma(\cdot)$ is
the Gamma function. Note that for positive integer $n$, $\Gamma(n) =
(n-1)!$. For the above type of density functions, the complementary
cumulative distribution function is given by \cite[Sec.
8.35]{Gradsh}
\begin{gather}
P\{z > x\} = 1 - \frac{1}{\Gamma(n)} \, \gamma(n, \lambda x) =
e^{-\lambda x} \sum_{m = 0}^{n-1} \frac{(\lambda x)^m}{m!}
\end{gather}
where $\gamma(\cdot,\cdot)$ is the incomplete Gamma function. Note
that $P\{z > x\}$ is a monotonically decreasing function of $x \ge
0$.

Next, we consider the function
\begin{gather}
f(x) = x p_z(x) = \frac{\lambda^n}{\Gamma(n)} x^n e^{-\lambda x}.
\end{gather}
It can be immediately found that the derivative of $f$ with respect
to $x$ is $\frac{d f(x)}{dx} = \frac{\lambda^n}{\Gamma(n)} x^{n-1}
e^{-\lambda x} (n -\lambda x)$, from which we conclude that $f$ is a
monotonically increasing function for $0 \le x \le n/\lambda$ and a
monotonically decreasing function for $x > n/\lambda$. Clearly, $f$
achieves its maximum at $x = n/\lambda$. Next, we show for all $x >
n/\lambda$ that
\begin{align} x p_z(x) - P\{z > x\} &= \frac{\lambda^n}{\Gamma(n)}
x^n e^{-\lambda x} - e^{-\lambda x} \sum_{m = 0}^{n-1}
\frac{(\lambda x)^m}{m!}
\\
&= e^{-\lambda x} \left( \frac{(\lambda x)^n }{(n-1)!} - \sum_{m =
0}^{n-1} \frac{(\lambda x)^m}{m!} \right) \label{eq:pdf-cdf0}
\\
&= e^{-\lambda x} \left( \frac{(\lambda x)^n}{(n-1)!} -\frac{
(\lambda x)^{n-1}}{(n-1)!} - \sum_{m = 0}^{n-2} \frac{(\lambda
x)^m}{m!} \right) \label{eq:pdf-cdf1}
\\
&= e^{-\lambda x} \left( \frac{(\lambda x)^{n-1} (\lambda x -
1)}{(n-1)!} - \sum_{m = 0}^{n-2} \frac{(\lambda x)^m}{m!} \right)
\label{eq:pdf-cdf2}
\\
&> e^{-\lambda x} \left( \frac{(\lambda x)^{n-1}}{(n-2)!} - \sum_{m
= 0}^{n-2} \frac{(\lambda x)^m}{m!} \right) \quad \forall x >
n/\lambda \label{eq:pdf-cdf3}
\\
&> 0 \quad \forall x > n/\lambda. \label{eq:pdf-cdf4}
\end{align}
Above, (\ref{eq:pdf-cdf1}) is obtained by writing $\sum_{m =
0}^{n-1} \frac{(\lambda x)^m}{m!} = \frac{ (\lambda
x)^{n-1}}{(n-1)!} - \sum_{m = 0}^{n-2} \frac{(\lambda x)^m}{m!}$.
(\ref{eq:pdf-cdf2}) follows after rearranging the terms.
(\ref{eq:pdf-cdf3}) is obtained by noting that $\lambda x - 1 > n -
1$ for $x > n/\lambda$, and therefore $\frac{\lambda x - 1}{(n-1)!}
> \frac{1}{(n-2)!}$. Finally, (\ref{eq:pdf-cdf4}) can easily be
verified by applying repetitively the same steps as in
(\ref{eq:pdf-cdf1}) -- (\ref{eq:pdf-cdf3}) to the other terms in the
summation $\sum_{m = 0}^{n-2} \frac{(\lambda x)^m}{m!}$.

(\ref{eq:pdf-cdf4}) shows that after reaching its maximum at $x =
n/\lambda$, the function $f$ is always greater than $P(z \geq x)$
and hence the two never intersect for $x > n/\lambda$. Note that,
for $0 \le x \le n/\lambda$, $f$ is a monotonically increasing
function. Moreover, $f(0) = 0$. On the other hand, $P\{z > x\}$
is always a monotonically decreasing function of $x \ge 0$. Note
also that at $x = 0$,  $P\{z > 0\} = 1$. Using these facts, we
immediately conclude that the function $f(x) = x p_z(x)$ and $P\{z
> x\}$ intersect only once in the interval $0 \le x \le
n/\lambda$. Therefore, $x p_z(x) = P(z > x)$ has a unique
solution for $x \ge 0$ when Gamma densities in the form given in
(\ref{appeq:gammadensity}) are considered.

\newpage

\begin{figure}
\begin{center}
\includegraphics[width=\figsize\textwidth]{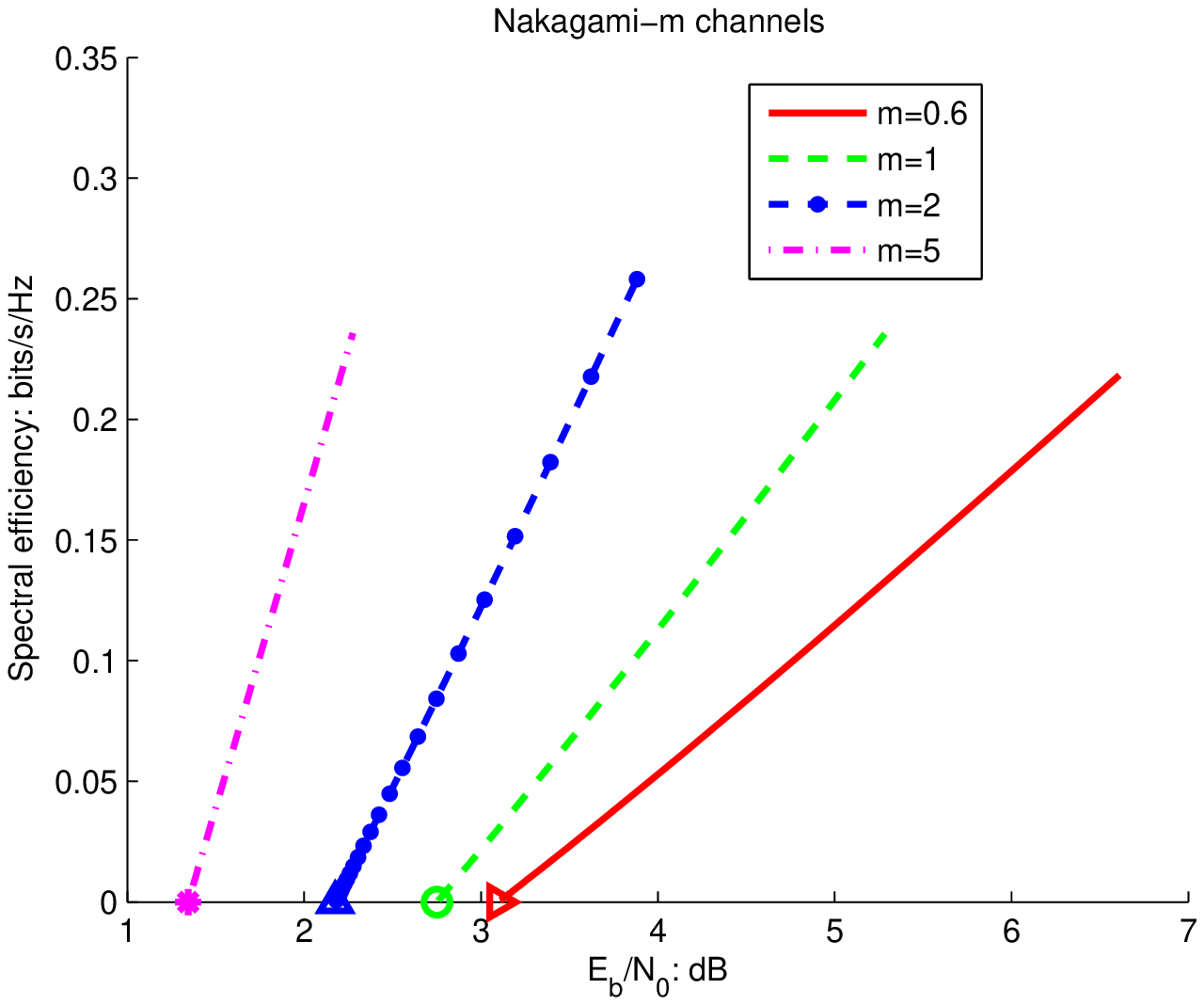}
\caption{Spectral efficiency vs. $E_b/N_0$ in Nakagami-$m$ channels;
$\theta=0.01$, $m = 0.6, 1, 2, 5$.}\label{fig:3}
\end{center}
\end{figure}

\end{document}